\documentclass[12pt]{article}

\usepackage{fancyhdr}
\usepackage{tocbibind}
\usepackage{amsmath}
\usepackage{amsbsy}
\usepackage{amssymb}
\usepackage{amscd}
\usepackage{amsfonts}
\usepackage{graphics}
\usepackage{verbatim}
\usepackage{subfigure}
\usepackage{xspace}
\usepackage{euscript}
\usepackage{alltt}
\usepackage{boxedminipage}
\usepackage{float}
\usepackage{color}
\usepackage{hyperref}
\usepackage[all]{xy}
\usepackage{t1enc}
\usepackage{times,exscale}
\usepackage{graphicx,calc}
\usepackage{mdwlist}
\usepackage{framed}
\usepackage{enumitem}
\usepackage{isomath}
\usepackage{amsmath}
\usepackage{amsbsy}
\usepackage{amssymb}
\usepackage{amscd}
\usepackage{amsfonts}

\newcommand{\R}{\mathbb R}

\newcommand{\Z}{\mathbb{Z}}

\newcommand{\bfa}{{\mathbf a}}
\newcommand{\bfb}{{\mathbf b}}
\newcommand{\bfc}{{\mathbf c}}

\newcommand{\bfe}{{\mathbf e}}
\newcommand{\bff}{{\mathbf f}}

\newcommand{\bfi}{{\mathbf i}}
\newcommand{\bfj}{{\mathbf j}}
\newcommand{\bfk}{{\mathbf k}}

\newcommand{\bfm}{{\mathbf m}}
\newcommand{\bfn}{{\mathbf n}}

\newcommand{\bfp}{{\mathbf p}}
\newcommand{\bfq}{{\mathbf q}}
\newcommand{\bfr}{{\mathbf r}}

\newcommand{\bfu}{{\mathbf u}}
\newcommand{\bfv}{{\mathbf v}}
\newcommand{\bfw}{{\mathbf w}}
\newcommand{\bfx}{{\mathbf x}}
\newcommand{\bfy}{{\mathbf y}}

\newcommand{\bfA}{{\mathbf A}}

\newcommand{\bfD}{{\mathbf D}}

\newcommand{\bfF}{{\mathbf F}}

\newcommand{\bfH}{{\mathbf H}}
\newcommand{\bfI}{{\mathbf I}}

\newcommand{\bfM}{{\mathbf M}}
\newcommand{\bfN}{{\mathbf N}}

\newcommand{\bfQ}{{\mathbf Q}}
\newcommand{\bfR}{{\mathbf R}}

\newcommand{\beq}{\begin{equation}}
\newcommand{\eeq}{\end{equation}}
\newcommand{\beqs}{\begin{eqnarray}}
\newcommand{\eeqs}{\end{eqnarray}}
\newcommand{\beql}{\begin{equation} \label}
\newcommand{\half}{\frac{1}{2}}

\newcommand{\bfzero}{\mathbf{0}}

\newcommand{\GCD}{\mathop{\rm GCD}\nolimits}

\newcommand{\xm}   {\mathbf x_{\bfzero,m}}
\newcommand{\xk}   {\mathbf x_{\bfzero,k}}
\newcommand{\xik}  {\mathbf x_{\bfi,k}}
\newcommand{\xjl}  {\mathbf x_{\bfj,l}}

\newcommand{\comments}[1]{}

\usepackage[margin=20mm]{geometry}
\numberwithin{equation}{section}

\newcommand{\titlescript}{Symmetry-Adapted Phonon Analysis of Nanotubes (to appear in J. Mech. Phys. Solids)}

\title{\titlescript}

\author{Amin Aghaei\footnote{Email: {\tt aghaei@cmu.edu}} , Kaushik Dayal\footnote{Email: {\tt kaushik@cmu.edu}} \ and Ryan S. Elliott\footnote{Email: {\tt elliott@aem.umn.edu}}
 \\ $^{* \dag}$ \small{Carnegie Mellon University, Pittsburgh, PA 15213}
 \\ $^\ddag$\small{University of Minnesota, Minneapolis, MN 55455}}
\date{\today}

\begin{document}
\pagestyle{fancyplain}
\lhead{\fancyplain{\scriptsize \titlescript
}
{\scriptsize \titlescript
}}
\rhead{\fancyplain{\scriptsize A. Aghaei, K. Dayal, R. S. Elliott}{\scriptsize A. Aghaei, K. Dayal, R. S. Elliott}}
\maketitle

\begin{center}
	{\bf Dedicated to Richard D. James on the occasion of his 60th Birthday}
\end{center}

\begin{abstract}
The characteristics of phonons, i.e. linearized normal modes of vibration, provide important insights into many aspects of crystals, e.g. stability and thermodynamics.
In this paper, we use the Objective Structures framework to make concrete analogies between crystalline phonons and normal modes of vibration in non-crystalline but highly symmetric nanostructures.
Our strategy is to use an intermediate linear transformation from real-space to an intermediate space in which the Hessian matrix of second derivatives is block-circulant.
The block-circulant nature of the Hessian enables us to then follow the procedure to obtain phonons in crystals: namely, we use the Discrete Fourier Transform from this intermediate space to obtain a block-diagonal matrix that is readily diagonalizable.
We formulate this for general Objective Structures and then apply it to study carbon nanotubes of various chiralities that are subjected to axial elongation and torsional deformation.
We compare the phonon spectra computed in the Objective Framework with spectra computed for armchair and zigzag nanotubes.
We also demonstrate the approach by computing the Density of States.
In addition to the computational efficiency afforded by Objective Structures in providing the transformations to almost-diagonalize the Hessian, the framework provides an important conceptual simplification to interpret the phonon curves.
Our findings include that, first, not all non-optic long-wavelength modes are zero energy and conversely not all zero energy modes are long-wavelength; second, the phonon curves accurately predict both the onset as well as the soft modes for instabilities such as torsional buckling; and third, unlike crystals where phonon stability does not provide information on stability with respect to non-rank-one deformation modes, phonon stability in nanotubes is sufficient to guarantee stability with respect to all perturbations that do not involve structural modes.
Our finding of characteristic oscillations in the phonon curves motivates a simple one-dimensional geometric nonlocal model of energy transport in generic Objective Structures.
The model shows the interesting interplay between energy transport along axial and helical directions.
\end{abstract}


\tableofcontents

\section{Introduction}

Phonons, i.e. normal modes, are extremely important to understand the properties of crystals.
For instance, phonon analysis provides insight into thermodynamic properties and mechanical stability \cite{dove-book,born-huang, elliott-jmps2006a,elliott-jmps2006b}.
In this paper, we use the framework of Objective Structures (OS) introduced by James \cite{james-jmps2006} to extend the notion of phonon analysis to noncrystalline but symmetric nanostructures.

OS generalizes the notion of a crystal or periodicity by using ideas from frame-indifference.
In brief, \cite{james-jmps2006} and following works have shown that many highly symmetric but non-crystalline nanostructures have close analogies to crystals.
These analogies have led to some important practical methods: e.g., a generalization of periodic boundary conditions for classical molecular dynamics and tight binding to enable the analysis of chiral nanostructures as well as the ability to apply torsional loads \cite{zhang2011helical,dumitrica-james-twist-1,dumitrica-james-twist-2,dumitrica-james-twist-3,DumitricaJames07,nikiforov-dumitrica,dayal-james-jmps2010,dayal-james-JFM2012,aghaei-dayal-msmse2012,Aghaei11}.
All of these exploit the fact that the symmetries of the nanostructure, together with frame-indifference, imply that the first derivative of the potential energy has certain symmetries.
This symmetry in the first derivative is a generalization of the fact that forces on image atoms are identical in a periodic crystal.

In this paper, we exploit in an essential way the symmetry in the second derivative of the potential energy, i.e. the Hessian matrix has various submatrices that are related to each other.
While this was noted by James \cite{james-jmps2006}, it has not been exploited in practical calculations.
Here, we find that this property implies that a preliminary linear transformation renders the Hessian matrix block-circulant as in periodic crystals, thus enabling the use of standard Fourier techniques after the preliminary transformation.
While a significant part of our analysis is general and applies broadly to all structures that belong to the family of OS, we also specialize the analysis to carbon nanotubes and other one-dimensional systems to do numerical calculations.
Except where stated as a model system, we use the well-characterized Tersoff interatomic potential for carbon that provides a balance between bond-order accuracy and computational efficiency \cite{Tersoff88}.

We emphasize that many researchers have studied phonons in carbon nanotubes for over a decade now.
For instance, an early example is \cite{kalia-1995-phonons}; a more recent review is \cite{dresselhaus-review2008}.
An important feature of all of these studies is that they directly use methods from periodic systems.
This often requires the use of very large unit cells that require expensive calculations to compute the phonon analysis, and more importantly the computed information is extremely complex and difficult to analyze for new physics.
Recent papers that exploit the symmetry of nanotubes are \cite{white2008lattice, popov-lambin}.

We go beyond these methods in some significant ways.
First, our approach based on OS provides a tight link to deformation of the nanotubes.
This is critical to go beyond exclusively axial-load-free and twisting-moment-free nanotubes; in fact, as recent work shows, the load- and moment- free structure of chiral nanotubes likely does {\em not} correspond to the assumed highly-symmetric configuration \cite{vercosa2010torsional, zhang2011helical,aghaei-dayal-msmse2012}.
Relaxing or applying such loads is readily accomplished using the OS framework \cite{james-jmps2006,DumitricaJames07}.

Second, our analysis exposes the close analogies to periodic crystals; additionally, it is general and applicable to a wide variety of OS that go beyond carbon nanotubes.
In addition to providing a conceptual unity, this can potentially enable important practical advances such as integrating our method with other techniques developed for periodic crystals.
For instance, we provide a demonstration of phonon soft-mode stability analysis to detect instabilities under torsion.
Such analyses, in combination with OS methods for bending and other deformations, can conceivably provide insight into complex phenomena such as nanotube rippling \cite{arroyo-3}.
Further, multiscale atomistic-to-continuum numerical methods have recently been used to predict structural phase transformations in crystals using phonon stability as a critical component \cite{QC-BFB-1,QC-BFB-2}; conceivably, similar methods for complex OS nanostructures can be built on the phonon approach provided here.
Another potentially important application of our phonon approach is, e.g., as a basis to construct effective Hamiltonian models that have been effective in predicting structural phase transformations in crystals based on soft modes.

The paper is organized as follows:
\begin{itemize}

 	\item Section \ref{notation} describes the notation used in the paper.

	\item Section \ref{OS-formulation} outlines the relevant aspects of OS, and the properties of the Hessian matrix in an OS.  In crystals, each row of the Hessian is simply a shifted copy of the previous row, i.e. it is block-circulant; in OS, each row of the Hessian is related to the previous row but in a more complex way that involves the symmetry parameters of the structure.

	\item Section \ref{phonons} sets up the linearized equations of motion, presents the transformation to the intermediate Objective space as a similarity transform, presents the standard Discrete Fourier Transform as a similarity transform, and uses a composition of these transforms to block-diagonalize the Hessian matrix.

	\item Section \ref{Example} shows examples of phonon spectra computed by the OS framework and contrasts these with standard phonon spectra.  The OS framework provides important conceptual simplifications in understanding these curves.  We also demonstrate the computation of the density of states.

	\item Section \ref{long-waves} examines modes that are long-wavelength (in the Fourier space), and rigid-body and uniform deformation modes in real space.  Unlike crystals, rigid-body modes in nanotubes are not always at long-wavelength, and long-wavelength non-optic modes do not always correspond to zero energy even in the limit.

	\item Section \ref{stability} examines the stability information provided by phonons, in particular contrasting nanotubes with crystals in regards to analogies to ``non-rank-one modes'' in crystals that are not tested by phonon stability.  We also demonstrate a numerical example of torsional buckling and examine the predictions of phonon soft-mode analysis.

	\item Section \ref{transport} studies energy transport in nanotubes.  We find characteristic oscillations in the phonon spectra that motivate a simple one-dimensional geometric nonlocal model of energy transport.  The model provides insight into the balance between energy transport along axial and helical directions.  The geometrically-motivated nature of the model gives it universal applicability to helical structures of all kinds.

\end{itemize}


\section{Notation}
\label{notation}
$\Z$ denotes the set of all integers and $\Z^3$  the set of triples of all integers.

For a quantity $\bfA$, the Fourier transform is denoted by $\tilde{\bfA}$.

To avoid ambiguity, the summation convention is {\em not} used and sums are always indicated explicitly.

Throughout the paper, $M$ and $N$ are used for the number of atoms per unit cell and the number of unit cells in the OS respectively.
Note that $M$ is always finite but $N$ can be infinite.
The unit cells are labeled by multi-indices denoted by boldface, i.e. $\bfi = (i_1, i_2, i_3)$, and atoms within a given unit cell are labeled by regular non-bold indices.
For example, the position of the $k$ atom in the $\bfi$ unit cell is denoted $\bfx_{\bfi,k}$.

Bold lower case and upper case letters represent vectors and matrices, respectively.
The rectangular Cartesian component and the exponent of a vector or matrix are shown respectively by Greek letters and lower case Latin letters as superscripts.

The subscripts of vectors and matrices are used to convey information about the structure of the quantity in addition to denoting components.
We will often deal with matrices of size $3MN\times 3MN$ corresponding to an OS with $3MN$ degrees of freedom (DoFs).
Such a matrix $\bfA$ can be divided into $N \times N$ blocks, with each block further sub-divided into $M \times M$ sub-blocks of size $3 \times 3$.
Then, $\bfA_{(\bfi,k)(\bfj,l)}$ denotes a $3\times 3$ sub-block, typically corresponding to the pair of atoms labeled by $(\bfi,k)$ and $(\bfj,l)$.
Also, $\bfA_{\bfi \bfj}$ denotes a $3M\times 3M$ matrix, typically corresponding to atoms in the unit cells labeled $\bfi$ and $\bfj$.

Similarly, for a vector $\bfb$ of size $3MN\times 1$, writing $\bfb_\bfi$ denotes the $\bfi$-th block of $\bfb$ of size $3M \times 1$ typically corresponding to the unit cell labeled by $\bfi$.
Further, $\bfb_{\bfi,k}$ denotes a sub-block of $\bfb_\bfi$ of size $3 \times 1$,  typically corresponding to the atom $k$ in the unit cell labeled by $\bfi$.

For Fourier quantities, the correspondences for sub-blocks in vectors and matrices are not to unit cells but rather to wave numbers.


\section{Objective Structures}
\label {OS-formulation}

James \cite{james-jmps2006} defined an objective atomic structure as a finite or infinite set of atoms in which every atom sees the same environment up to translation and rotation.
Similarly, an objective molecular structure is defined as a structure with a number of identical {\em molecules}, each molecule consisting of a number of atoms, arranged such that {\em corresponding} atoms in every molecule see the same environment up to translation and rotation.
We note that the molecules in an objective molecular structure need not correspond to standard physical molecules as usually understood.
Bravais (multi) lattices are special cases of objective atomic (molecular) structures in which each atom (molecule) has the same environment up to translation and the rotation is trivial.

Following recent works that build on James' original formulation, e.g. \cite{dayal-elliott-james-formulas, dayal-james-JFM2012, dayal-james-jmps2010, Aghaei11, aghaei-dayal-msmse2012}, we can define OS equivalently in the language of group theory.
The group theoretic approach enables practical calculations.
Let $G=\{ g_{\bf 0},g_{\bf 1},\cdots,g_{\bfN} \}$ be a set of isometries indexed by a multi-index.
Each element of $G$ has the form $g_{\bfj}=( \bfQ_{\bfj} | \bfc_{\bfj} )$ where $\bfQ_{\bfj} \in O(3)$ is orthogonal and $\bfc_{\bfj} \in \R^3$ is a vector.

The action of an isometry on a point $\bfx \in \R^3$ is
\begin{equation}
	g_{\bfj} (\bfx) = \bfQ_{\bfj} \bfx +\bfc_{\bfj}
\end{equation}
Composition of mappings then provides:
\begin{equation*}
	g_\bfi (g_\bfj (\bfx)) = \bfQ_\bfi \left( \bfQ_\bfj \bfx + \bfc_\bfj \right) + \bfc_\bfi = \bfQ_\bfi \bfQ_\bfj \bfx + \bfQ_\bfi \bfc_\bfj + \bfc_\bfi
\end{equation*}
This motivates a definition for multiplication of isometries:
\begin{equation}
 \label{multiplic}
	g_\bfi g_\bfj  =  (\bfQ_\bfi \bfQ_\bfj | \bfQ_\bfi \bfc_\bfj + \bfc_\bfi) 
\end{equation}
From this definition, it follows that the identity element is $g_{\bf 0}:=(\bfI | \bfzero)$ and the inverse of $g_\bfi$ is defined by $g_\bfi^{-1} := (\bfQ_\bfi^T | -\bfQ_\bfi^T \bfc_\bfi)$.

If the set $G$ is additionally a group with respect to the multiplication operation above, then placing an atom at each of the points given by the action of elements of $G$ on a given point $\bfx_0$ gives an objective atomic structure.
In addition, placing an atom of species $k$ at each of the points given by the action of elements of $G$ on a given set of points $\bfx_{0, k}$ gives an objective molecular structure.

In this paper, we will consider OS described by groups of the form
\begin{equation}
	G=\{ g_1^{i_1} g_2^{i_2} g_3^{i_3} \; ; \; (i_1, i_2, i_3) \in \Z^3 \}
\end{equation}
Here, $g_1=(\bar\bfQ_1 | \bar\bfc_1)$, $g_2=(\bar\bfQ_2 | \bar\bfc_2)$ and  $g_3=(\bar\bfQ_3 | \bar\bfc_3)$ are the {\em generators} of the group.
We assume that they commute, i.e. $g_1 g_2 = g_2 g_1$ and so on.
This immediately implies that $G$ itself is Abelian.
As shown in \cite{dayal-elliott-james-formulas}, $G$ of this form does not describe all possible OS; however, as also shown there, those OS that cannot directly be described can nevertheless be described by such a $G$ by enlarging the unit cell and neglecting certain intra-unit cell symmetries.

Denoting the atomic positions in the unit cell by $\bfx_{(0,0,0),k}:= \xk, k=1,\ldots,M$, the OS is described by
\begin{equation} \label{obj}
	\xik := \bfx_{(i_1,i_2,i_3),k}=g_1^{i_1} g_2^{i_2} g_3^{i_3} (\xk) = \bfQ_\bfi \xk +\bfc_\bfi \quad ; \quad \bfi=(i_1,i_2,i_3)\in \Z^3
\end{equation}
Using \eqref{multiplic}, we have that
\begin{equation} \label{obj2}
	\bfQ_\bfi = \bar\bfQ_1^{i_1} \bar\bfQ_2^{i_2} \bar\bfQ_3^{i_3}, \quad
	\bfc_\bfi =\bar\bfQ_1^{i_1} \bar\bfQ_2^{i_2} \left( \sum_{p=0}^{p=i_3-1} \bar\bfQ_3^p \bar\bfc_3 \right)
			+ \bar\bfQ_1^{i_1} \left( \sum_{p=0}^{p=i_2-1} \bar\bfQ_2^p \bar\bfc_2 \right)
			+ \left( \sum_{p=0}^{p=i_1-1} \bar\bfQ_1^p \bar\bfc_1 \right)
\end{equation}
for positive exponents $i_1, i_2, i_3$.  Negative exponents are defined through the inverse.

An important property of OS is that elements of $G$ map images of the unit cell $\xk$ to each other.
For instance, consider the $\bfi$ and $\bfj$ images of the unit cell:
\begin{equation} \label{obj3}
	\xik
	=  g_1^{i_1} g_2^{i_2} g_3^{i_3} (\xk)
	=  \left(g_1^{i_1-j_1} g_2^{i_2-j_2} g_3^{i_3-j_3} \right) \left( g_1^{j_1} g_2^{j_2} g_3^{j_3} \right) (\xk)
	=  \left(g_1^{i_1-j_1} g_2^{i_2-j_2} g_3^{i_3-j_3} \right) (\bfx_{\bfj, k})
\end{equation}
These relations follow directly from the closure and commuting properties of the Abelian group $G$.

Consider the orthogonal part of $g_1^{i_1-j_1} g_2^{i_2-j_2} g_3^{i_3-j_3}$:
\begin{equation*}
	(\bfQ_{\bfi-\bfj}|\bfc_{\bfi-\bfj})
	 = g_1^{i_1-j_1} g_2^{i_2-j_2} g_3^{i_3-j_3}
	 = \left( g_1^{i_1} g_2^{i_2} g_3^{i_3} \right) \left( g_1^{-j_1} g_2^{-j_2} g_3^{-j_3} \right)
	 = (\bfQ_\bfi|\bfc_\bfi) (\bfQ_\bfj|\bfc_\bfj)^{-1}
	 = (\bfQ_\bfi \bfQ_\bfj^T | -\bfQ_\bfi \bfQ_\bfj^T \bfc_\bfj + \bfc_\bfi )
\end{equation*}
This provides the important relation:
\begin{equation} \label{Qtranspos}
	\bfQ_\bfi \bfQ_\bfj^T = \bfQ_{\bfi-\bfj}
\end{equation}
%
%


\subsection{Crystal Lattices and Carbon Nanotubes as Objective Structures} \label{Def:SWNT_OS}

Two important examples of OS are crystal multilattices and carbon nanotubes.  We describe them using the general OS framework above.

To describe a crystal multilattice as an OS, we simply set $\bar\bfQ_1=\bfI,\bar\bfQ_2=\bfI,\bar\bfQ_3=\bfI$.
The vectors $\bar\bfc_1, \bar\bfc_2, \bar\bfc_3$ are the lattice vectors.

Carbon nanotubes require only two generators.
Therefore, we set $g_3$ to the identity.
For a nanotube with axis $\bfe$ and centered at $\bf 0$, we use: 
\begin{equation}
	 \label{NT_g1g2-1}
	g_1 = (\bfR_{\theta_1}|\mathbf 0), \bfR_{\theta_1} \bfe = \bfe; \qquad g_2 =(\bfR_{\theta_2}|\kappa_2 \bfe), \quad \bfR_{\theta_2}\bfe = \bfe
\end{equation}
following (\ref{NT_g1g2}).
Here $\bfR_\theta$ is a rotation tensor with angle $\theta$.
The generator $g_1$ is a rotation isometry, and $g_2$ is a screw isometry
The parameters $\kappa_2, \theta_1, \theta_2$ depend on the chiral indices $(m,n)$ of the nanotube.
In this description, the unit cell has 2 atoms at positions $\bfx_{(0,0),0}$ and $\bfx_{(0,0),1}$.
The relations between these parameters and $(m,n)$ are described in Appendix \ref{SWNT-parameters}.

We note the important special case that when the chiral indices $m$ and $n$ of the nanotube are relatively prime, $g_1$ reduces to the identity.

The 2-atom unit cell of (\ref{NT_g1g2-1}) is sufficient to obtain the dispersion curves of carbon nanotubes.
But, as in standard periodic calculations, small unit cells can also greatly constrain the possible deformations.
This is of particular concern when using (zero temperature) atomistics to study instabilities that lead to defects.
For such problems, a unit cell with more atoms can be useful.
To this end, we first define an enlarged unit cell consisting of the atoms generated by $g_1^p g_2^q (\bfx_{(0,0),k})$, where $k=1,2$ and the indices $p$ and $q$ run over integers $p_1 \le p \le p_2$ and $q_1 \le q \le q_2$.
The unit cell now consists of $2(p_2-p_1+1)(q_2-q_1+1)$ atoms.
To generate the nanotube, we then define the group $H=\{ h_1^i h_2^j ; (i,j)\in \Z^2\}$ with $h_1= g_1^{p_2 - p_1 +1}$ and $h_1= g_1^{p_2 - p_1 +1}$.
Note that $H$ is a subgroup of $G$.
The action of elements of $H$ on the enlarged unit cell define precisely the same nanotube as using $G$ on the 2-atom unit cell.
However, since the atoms in the enlarged unit cell are not constrained to each other by symmetry, they can explore a larger space of deformations.


\subsection{Consequences of Frame Indifference on the Potential Energy and its Derivatives}

By frame-indifference, the potential energy of the OS, $\phi(\bfx_{\bfzero,k},\ldots,\bfx_{\bfi,k},\ldots)$ where $k=1,\ldots,M$ and $\bfi=(i_1,i_2,i_3)\in \Z^3$, is invariant under rigid translations and rotations of the entire structure, assuming that external fields are either absent or also similarly transform.
We apply the specific rigid translation and rotation associated to elements of $G$, i.e., consider the transformation $ g_{\bf -i} : =  g_1^{-i_1} g_2^{-i_2} g_3^{-i_3}$.
\begin{equation}\label{pot}
	\phi \left(\xk,\ldots,\xik,\ldots,\xjl,\ldots \right)
	= \phi \left(g_{\bf -i} \xk,\ldots,g_{\bf -i} \xik,\ldots,g_{\bf -i}\xjl,\ldots \right)
	= \phi \left(\bfx_{-\bfi,k},\ldots,\xk,\ldots,\bfx_{\bfj-\bfi,l},\ldots \right)
\end{equation}

The key observation that enabled Objective Molecular Dynamics \cite{DumitricaJames07, dayal-james-jmps2010} is as follows.
The force on atom $(\bfi,k)$ is $\bff_{\bfi,k}=-\frac{\partial \phi}{\partial \xik}$.
Starting from the potential energy in (\ref{pot}), we perturb atom $(\bfi,k)$ along the coordinate direction $\bfe^\alpha$ and atom $(\bfj,l)$ along the coordinate direction $\bfe^\beta$.
Formally, we can write:
\begin{align} \label{PotDer}
	\phi & \left(\xm,\ldots,\xik+\epsilon_1 \bfe^\alpha,\ldots,\xjl+\epsilon_2 \bfe^\beta,\ldots \right) \notag \\
	& = \phi \left(g_{-\bfi} \xm,\ldots,g_{-\bfi} (\xik+\epsilon_1 \bfe^\alpha),\ldots,g_{-\bfi} (\xjl+\epsilon_2 \bfe^\beta),\ldots \right) \notag \\
	& = \phi \left(\bfx_{-\bfi,m},\ldots,\xk+\epsilon_1 \bfQ_\bfi^T \bfe^\alpha,\ldots,\bfx_{\bfj-\bfi ,l}+\epsilon_2 \bfQ_\bfi^T \bfe^\beta,\ldots \right) \notag \\
	& = \phi \left(\xk+\epsilon_1 \bfQ_\bfi^T \bfe^\alpha,\ldots,\bfx_{\bfn,m},\ldots,\bfx_{\bfj-\bfi ,l}+\epsilon_2 \bfQ_\bfi^T \bfe^\beta,\ldots \right)
\end{align}
The calculation is justified as follows: from the first to the third line, we follow precisely (\ref{pot}), and in the last step we simply rearrange the arguments because the energy does not not depend on the labeling of the atoms.

Setting $\epsilon_2=0$ identically and taking the limit of $\epsilon_1 \rightarrow 0$:
\begin{equation} \label{force1}
	\frac{\partial \phi}{\partial x_{\bfi,k}^\alpha}= \sum_{\gamma = 1}^{3} Q_\bfi^{\alpha \gamma} \frac{\partial \phi}{\partial x_{\bfzero,k}^\gamma}
	 \Leftrightarrow  \bff_{\bfi,k}= \bfQ_\bfi \bff_{\bfzero,k}
\end{equation}
This transformation law for the force acting on an atom in the unit cell and its images enables the analog of periodic molecular dynamics in general OS \cite{DumitricaJames07, dayal-james-jmps2010}.

James \cite{james-jmps2006} also noted a similar transformation law for elements of the second derivative (Hessian) matrix $\bfH$.
Taking the limit consecutively of $\epsilon_1 \rightarrow 0$ and $\epsilon_2 \rightarrow 0$:
\begin{equation} \label{hessian1}
	\frac{\partial^2 \phi}{\partial x_{\bfi,k}^\alpha \partial x_{\bfj,l}^\beta}
	=\sum_{\gamma, \eta} Q_\bfi^{\alpha \gamma} \frac{\partial^2 \phi}{\partial x_{\bfzero,k}^\gamma \partial x_{\bfj-\bfi,l}^\eta}  Q_\bfi^{\beta \eta}
	\Leftrightarrow \bfH_{(\bfi,k)(\bfj,l)}= \bfQ_\bfi \bfH_{(\bfzero,k)(\bfj-\bfi,l)} \bfQ_\bfi^T
\end{equation}
Note that for a periodic crystal $\bfQ_\bfi = \bfI$ for all $\bfi$, therefore providing the standard relation that $\bfH$ is block circulant.

The key physical content of (\ref{hessian1}) is that interactions between any pair of atoms in the structure can be mapped to the interactions between atoms in the unit cell and atoms in some other image cell.


\section{Normal Mode Analysis in Objective Structures} \label{phonons}

We first derive the standard linearized equations of motion in an OS.
We then use the properties of the Hessian from the previous section to achieve a block-diagonalization of the Hessian such that each block is of size $3M \times 3M$.
Each block is related to the frequency in Fourier space, but the transformation is not directly from real to Fourier space but goes through an intermediate linear transform.

\subsection{Linearized Equation of Motion around an Equilibrium Configuration}

Let $\mathring\bfx=\{\mathring\bfx_{\bfi,k} \; ; k=1,\ldots,M\; ; \bfi\in \Z^3 \}$ be the equilibrium configuration.
Consider a perturbation $\bfu_{\bfi,k}$ about this configuration and use a Taylor expansion:
\begin{equation}
	\phi(\mathring\bfx + \bfu)
	= \phi(\mathring\bfx)
	+ \sum_{(\bfi,k)} \sum_{\alpha=1}^3 \frac{\partial \phi}{\partial x_{\bfi,k}^\alpha} \bigg|_{\mathring\bfx} u_{\bfi,k}^\alpha
	+ \half \sum_{(\bfi,k)} \sum_{\alpha=1}^3 \sum_{(\bfj,l)} \sum_{\beta=1}^3 \frac{\partial^2 \phi}{\partial x_{\bfi,k}^\alpha \partial x_{\bfj,l}^\beta} \bigg|_{\mathring\bfx} u_{\bfi,k}^\alpha u_{\bfj,l}^\beta + \ldots
\end{equation}
Since $\mathring\bfx$ is an equilibrium configuration, the first derivative does not appear.
Neglecting terms higher than quadratic:
\begin{equation}
	\phi(\mathring\bfx + \bfu)= \phi(\mathring\bfx) + \half \sum_{(\bfi,k)} \sum_{\alpha=1}^3 \sum_{(\bfj,l)} \sum_{\beta=1}^3 \frac{\partial^2 \phi}{\partial x_{\bfi,k}^\alpha \partial x_{\bfj,l}^\beta} \bigg|_{\mathring\bfx} u_{\bfi,k}^\alpha u_{\bfj,l}^\beta
\end{equation}
Taking the limit of $\bfu \rightarrow {\bf 0}$, the force on the atom $(\bfi,k)$ is:
\begin{equation} \label{PhForce1}
	m_k \ddot{u}^{\alpha}_{\bfi,k}
	= f_{\bfi,k}^\alpha
	:= - \frac{\partial \phi}{\partial x_{\bfi,k}^\alpha}
	= -\sum_{(\bfj,l)} \sum_{\beta=1}^3 \frac{\partial^2 \phi}{\partial x_{\bfi,k}^\alpha \partial x_{\bfj,l}^\beta} u_{\bfj,l}^\beta
	= -\sum_{(\bfj,l)} \sum_{\beta=1}^3 H_{(\bfi,k)(\bfj,l)}^{\alpha\beta} u_{\bfj,l}^\beta
\end{equation}
In compact matrix form,
\begin{equation} \label{EqMotion2}
	\bfM \ddot\bfu = -\bfH \bfu
\end{equation}
$\bfM$ and $\bfH$ are the $3MN\times 3MN$ mass and Hessian matrices respectively.
Note that $\bfM$ is diagonal and trivially inverted.
So define $\hat\bfH:=\bfM^{-1}\bfH$.

The linear form of (\ref{EqMotion2}) implies that solutions are exponentials, i.e., $\bfu = \hat\bfu \exp(-i\omega t)$ where $\omega$ is the angular frequency.
Therefore, we seek to solve the eigenvalue problem:
\begin{equation} \label{EqMotion4}
	\omega^2_p \hat\bfu^p = \hat \bfH \hat\bfu^p
\end{equation}
$\omega^2_p$ and $\hat\bfu^p$ are the eigenvalues and eigenvectors of $\hat\bfH$, with $p=1,\ldots,3MN$.
In generic finite structures, solving this eigenvalue problem can be computationally demanding.
In infinite periodic crystals, $\hat\bfH$ is block-circulant as noted above.
Consequently, it can be block-diagonalized using the Fourier transform, thus converting the problem of solving a $3MN$ system into solving $N$ systems of size $3M$.
Note that this is only formally true, because $N$ is infinite in this case; in addition, the Fourier transform enables more than just computational saving as it provides important physical insights to organize the nominally infinite number of solutions.
Therefore, instead of finding the eigenvalues of a $3MN\times 3MN$ matrix, we can calculate the eigenvalues of $3M \times 3M$ matrices $N$ times (Appendix \ref{DFT}).
For an OS however, $\hat\bfH$ is not block-circulant.  We deal with this in Section \ref{DiagOS}.



\subsection{Block-diagonalization for an Objective Structure} \label{DiagOS}

As noted immediately above, the Hessian $\hat\bfH$ in an OS is not block-circulant; however, there is a close analogy in (\ref{hessian1}), i.e. $\hat\bfH_{(\bfp,k)(\bfq,l)}= \bfQ_\bfp \hat\bfH_{(\bfzero,k)(\bfq-\bfp,l)} \bfQ_\bfp^T$.
As we show below, the linear transformation defined by the $3MN \times 3MN$ matrix
\begin{equation} \label{OStransf1}
	\bfR_{(\bfp,k)(\bfq,l)}= \bfQ_\bfp \delta_{\bfp\bfq} \delta_{kl}
\end{equation}
takes us to the {\em Objective Space} in which $\hat\bfH$ is block-circulant.
It is then possible to block-diagonalize $\hat\bfH$ using the DFT.

In the one-dimensional case $\bfR$ has the form:
\begin{equation} \label{bigR2}
	\bfR=
		\begin{bmatrix}
			[\bfR_{00}] & [\bfzero]   & \cdots & [\bfzero]     \\
			[\bfzero]   & [\bfR_{11}] & \cdots & [\bfzero]     \\
			\vdots      & \vdots      & \ddots & \vdots        \\
			[\bfzero]   & [\bfzero]   & \cdots & [\bfR_{(N-1)(N-1)}] 
		\end{bmatrix}
\end{equation}
We note that one-dimensional does not refer to real-space but rather to the number of slots in the multi-index that indexes the unit cells.

Each submatrix $\bfR_{pp}$ is a $3M\times 3M$ block-diagonal matrix
\begin{equation} \label{bigR3}
	\bfR_{pp}=
		\begin{bmatrix}
			[\bfQ_p]  & [\bfzero] & \cdots & [\bfzero]   \\
			[\bfzero] & [\bfQ_p]  & \cdots & [\bfzero]   \\
			\vdots    & \vdots    & \ddots & \vdots      \\
			[\bfzero] & [\bfzero] & \cdots & [\bfQ_p]
		\end{bmatrix}
\end{equation}
and each $\bfQ_j$ is a $3\times 3$ orthogonal tensor. $\bfR$ is obviously orthogonal.

Defining $\hat\bfu=\bfR \hat\bfv$ and substituting this in (\ref{EqMotion4}), we get 
\begin{equation} \label{OSeigen1}
	\omega^2 \bfR \hat\bfv = \hat \bfH  \bfR \hat\bfv
	\Rightarrow
	\omega^2 \hat\bfv = \bfR^T \hat\bfH \bfR \hat\bfv = \hat\bfD \hat\bfv
\end{equation}
where $\hat\bfD := \bfR^T \hat\bfH \bfR$ is the transformed Hessian matrix.
We now show that $\hat\bfD$ is block-circulant.
Using (\ref{OStransf1}), we have that:
\begin{equation} \label{Dynamical1}
	\hat\bfD_{(\bfp,k)(\bfq,l)}= \bfQ_\bfp^T \hat\bfH_{(\bfp,k)(\bfq,l)} \bfQ_\bfq
\end{equation}
Now, substitute (\ref{hessian1}) into (\ref{Dynamical1}):
\begin{align} \label{Dynamical2}
	\hat\bfD_{(\bfp,k)(\bfq,l)}
	& = \bfQ_\bfp^T \bfQ_\bfp \hat\bfH_{(\bfzero,k)(\bfq-\bfp,l)} \bfQ_\bfp^T \bfQ_\bfq \notag \\
	& = \bfQ_\bfzero^T \hat\bfH_{(\bfzero,k)(\bfq-\bfp,l)} \bfQ_{\bfq-\bfp} \notag \\
	& = \hat\bfD_{(\bfzero,k)(\bfq-\bfp,l)}
\end{align}
where we have used (\ref{Qtranspos}) and $\bfQ_\bfzero=\bfI$.

Therefore, $\hat\bfD$ is block-circulant and can be block-diagonalized by the DFT as described in Appendix \ref{DFT}.
Essentially, we use two successive linear transforms to solve
\begin{equation} \label{OSeigen3}
	\omega^2 \tilde\bfv = \tilde\bfD \tilde\bfv
\end{equation}
where $\tilde\bfD$ is block-diagonal and
\begin{subequations} \begin{align}
	\tilde\bfv = & \bfF \hat\bfv = \bfF \bfR^T \hat\bfu   \label{Dynamical3}  \\
	\tilde\bfD = & \bfF \hat\bfD \bfF^{-1} = \bfF \bfR^T \hat\bfH \bfR \bfF^{-1} = \bfF \bfR^T \bfM^{-1} \bfH \bfR \bfF^{-1} \label{Dynamical4}
\end{align} \end{subequations}

Since $\tilde\bfD$ is block-diagonal, we can simplify the dynamical equation \eqref{OSeigen3} to read:
\begin{equation} \label{OSeigen4}
	\left(\omega^2 \right)^{[\bfp]} \tilde\bfv_\bfp = \tilde\bfD_{\bfp\bfp} \tilde\bfv_\bfp
\end{equation}
where $\left(\omega^2 \right)^{[\bfp]}$ is the eigenvalue corresponding to the eigenvector $\tilde\bfv_\bfp$.
From \eqref{BlockDiag} and \eqref{Dynamical2}
\begin{align} \label{OSeigen5}
	\tilde\bfD_{\bfp\bfp}
	& = \sum_\bfr \exp \big[-i \bfk_\bfp \cdot \bfy_\bfr \big] \hat\bfD_\mathbf{0r}  \notag \\
	& = \sum_\bfr \exp \big[-i \bfk_\bfp \cdot \bfy_\bfr \big] \hat\bfH_\mathbf{0r} \bfR_{\bfr\bfr} 
\end{align}
For the wave vector associated with $\bfp$, \eqref{OSeigen4} gives $3M$ solutions analogous to the multiple branches in a phonon spectrum.
We index these by $\nu$.
The displacement of the atom $(\bfq,l)$ induced by the normal mode labeled by the wave-vector $\bfp$ and the $\nu$-th branch is obtained by solving for $\hat \bfu$ in  \eqref{Dynamical3} and using \eqref{eqn:Fourier-matrix} and \eqref{OStransf1}:
\begin{equation}\label{Polariz2}
	\hat\bfu_{(\bfq,l)}^{[\bfp,\nu]}= \frac{1}{\sqrt{N}} \bfQ_\bfq \tilde\bfv_{(\bfp,l)}^{[\nu]} \exp \big[ -i \bfk_\bfp \cdot \bfy_\bfq \big]
\end{equation}

As can be seen from \eqref{Dynamical4}, $\bfR$ acts first on the Hessian matrix $\hat\bfH$ and ``unwraps'' the structure by transforming it to Objective Space.
Subsequently, $\bfF$ acts on the unwrapped periodic structure.
Therefore both position vector $\bfy$ and wave vector $\bfk$ are defined in Objective Space.
The quantities $\bfk_\bfp \cdot \bfy_\bfq$ can be obtained using the standard method for periodic systems outlined in Appendix \ref{DFT}.

We summarize the key steps in our algorithm:
\begin{framed}
	\begin{enumerate}[topsep=0pt, partopsep=0pt, itemsep=0pt]
		\item Calculate $\hat \bfH_\mathbf{0r}$, the Hessian matrix corresponding to interactions between any chosen unit cell, labeled ${\bf 0}$, and the neighboring cells.   The size of $\hat \bfH_\mathbf{0r}$ is $3M \times 3MN$.
		\item Multiply each vertical block of $\hat \bfH_\mathbf{0r}$ by the rotation matrix of that block, i.e. calculate $\hat \bfH_\mathbf{0r} \bfR_{\bfr\bfr}$.
		\item Calculate $\tilde\bfD_{\bfp\bfp}= \sum_\bfr \exp \big[-i \bfk_\bfp \cdot \bfy_\bfr \big] \hat\bfH_\mathbf{0r} \bfR_{\bfr\bfr}$, the dynamical matrix associated with wave vector $\bfk_\bfp$.
		\item Find the eigenvalues, $\left( \omega^2 \right)^{[\bfp,\nu]}$, and eigenvectors, $\tilde \bfv_\bfp^{[\nu]}$, of $\tilde\bfD_{\bfp\bfp}$. 
		\item The normalized displacement of the atoms are $\hat\bfu_{(\bfq,l)}^{[\bfp,\nu]}= \frac{1}{\sqrt{N}} \bfQ_\bfq \tilde\bfv_{(\bfp,l)}^{[\nu]} \exp \big[ -i \bfk_\bfp \cdot \bfy_\bfq \big]$.  The wave vector and the branch number respectively are labeled by $\bfp$ and $\nu$.
	\end{enumerate}
\end{framed}


\section{A Numerical Example: Dispersion Curves of $(6,6)$ Carbon Nanotubes} \label{Example}

While the OS framework can be used for nanotubes with any chirality, in this section we focus on $(6,6)$ carbon nanotubes to illustrate some typical features of the phonon curves.
This particular chirality also has a small translational unit cell that enables comparisons with standard periodic calculations.

\begin{figure}[htbp]
	\centering
	\subfigure{\footnotesize (a)}\includegraphics[trim=10mm 70mm 0mm 70mm, height=10.8cm,clip]{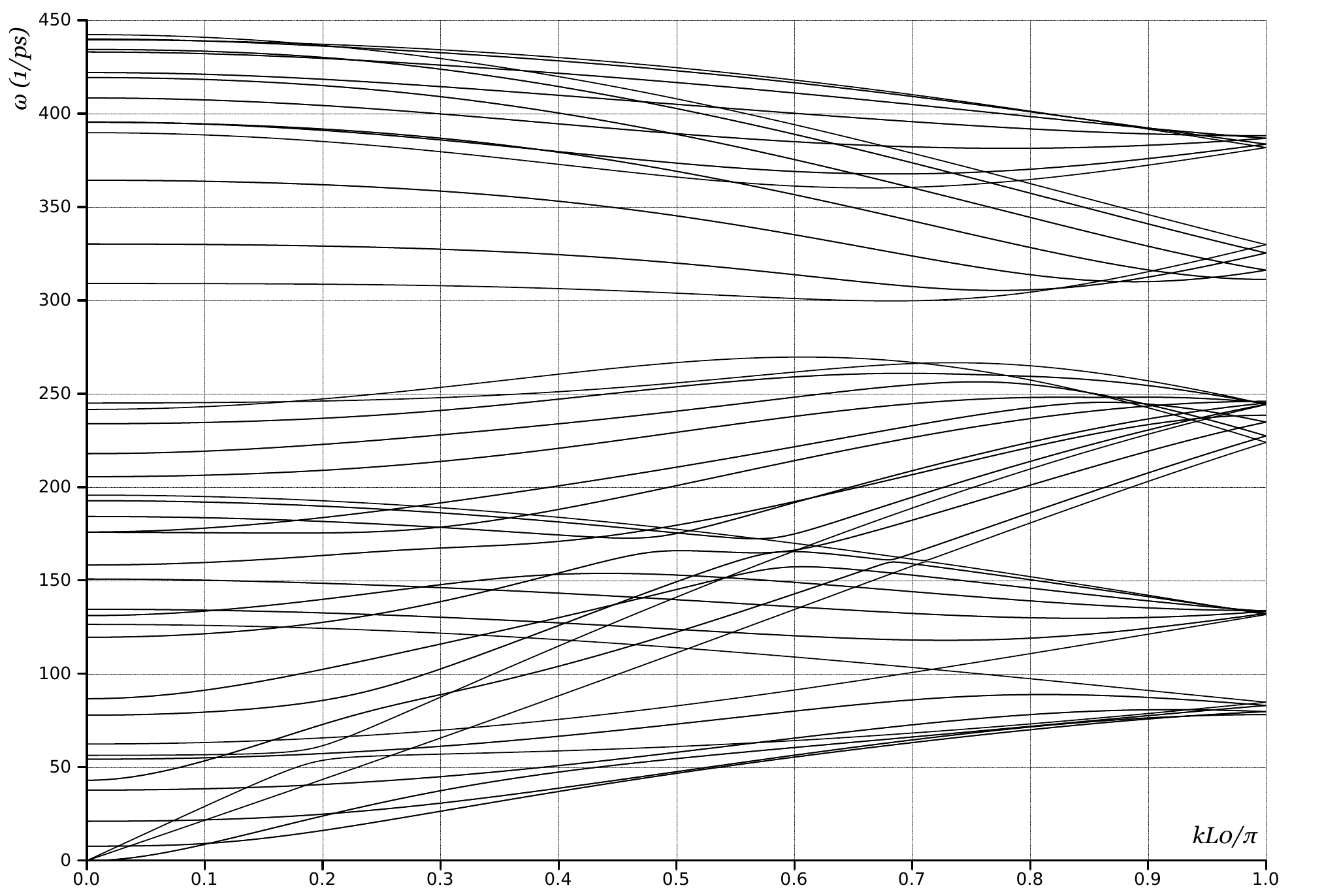}
	\subfigure{\footnotesize (b)}\includegraphics[trim=10mm 70mm 0mm 70mm, height=10.8cm,clip]{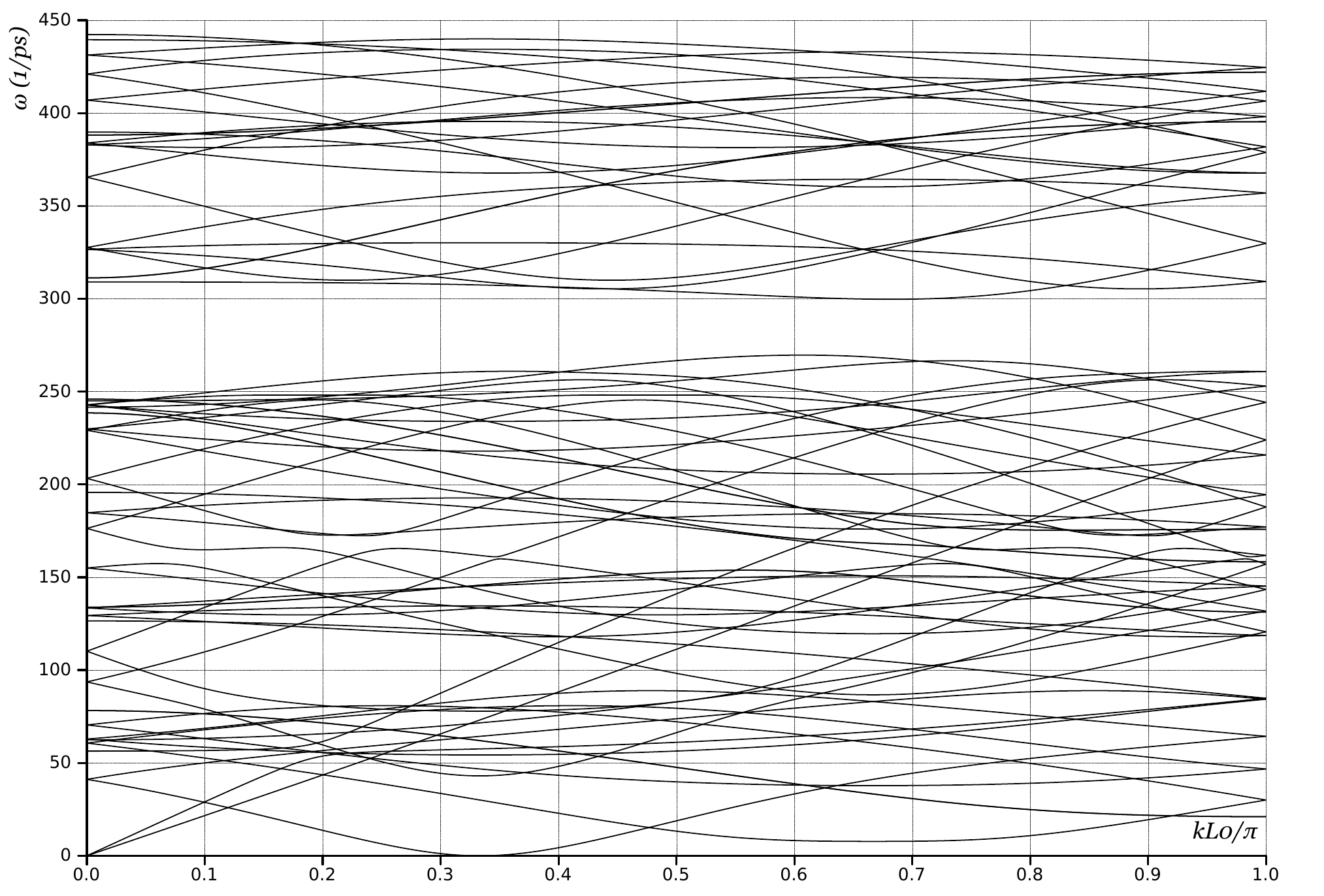}
	\caption{Dispersion curves of a $(6,6)$ carbon nanotube for (a) Choice 1 and (b) Choice 2.  The wavevector is normalized by the length of the translation vector $0.246 \text{nm}$ in $g_2$.  The large number of phonon curves that have equal and opposite slopes at the right edge of the plots (i.e. ``folded over'') are a signature of the large unit cell.}
	\label{fig:n6m6_24A}
\end{figure}

\begin{figure}[htbp]
	\centering
	\includegraphics[trim=0mm 70mm 0mm 70mm, height=11cm, clip]{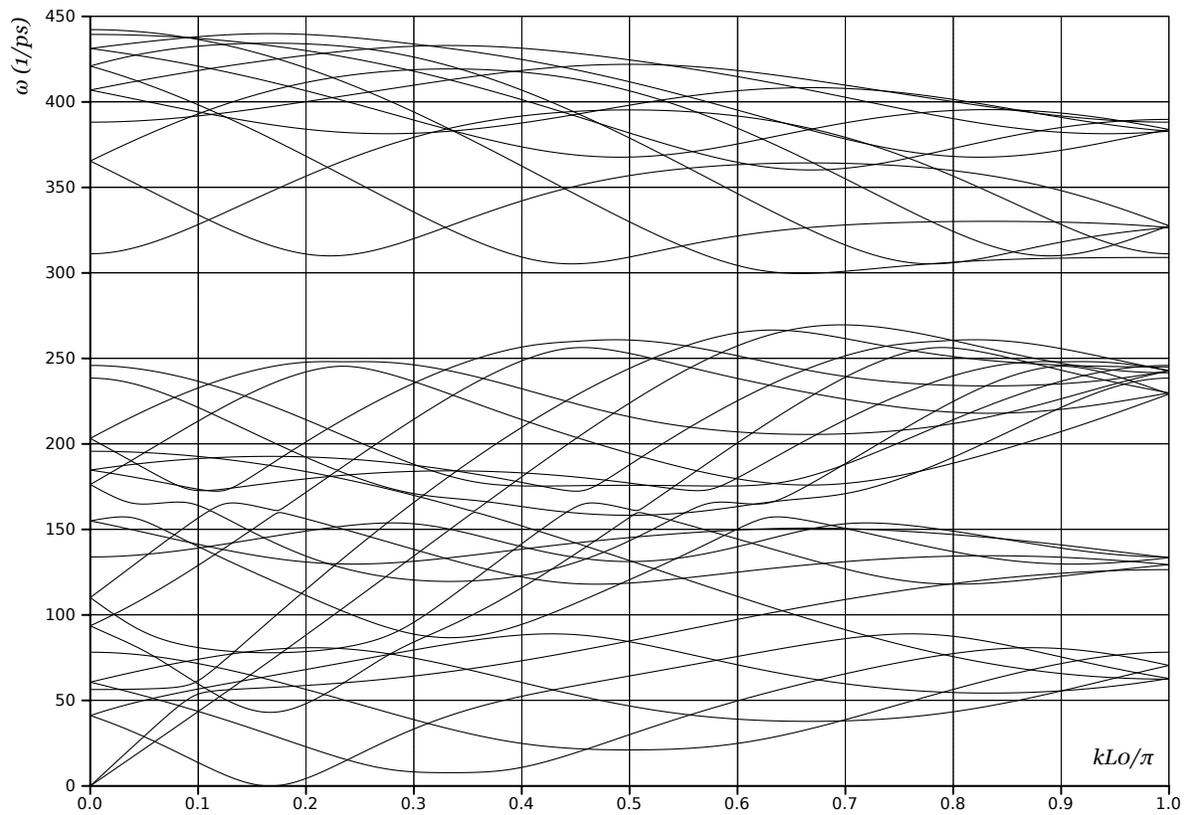}
	\caption{Dispersion curves of a $(6,6)$ carbon nanotubes using Choice 3.  The wavevector is normalized by the length of the translation vector $0.123 \text{nm}$ in $g_2$.  The band-folding shows that our unit cell still has unused symmetries.}
	\label{fig:n6m6_12A}
\end{figure}

\begin{figure}[htbp]
	\centering
	\includegraphics[trim=14mm 35mm 7mm 35mm, height=18.5cm, clip]{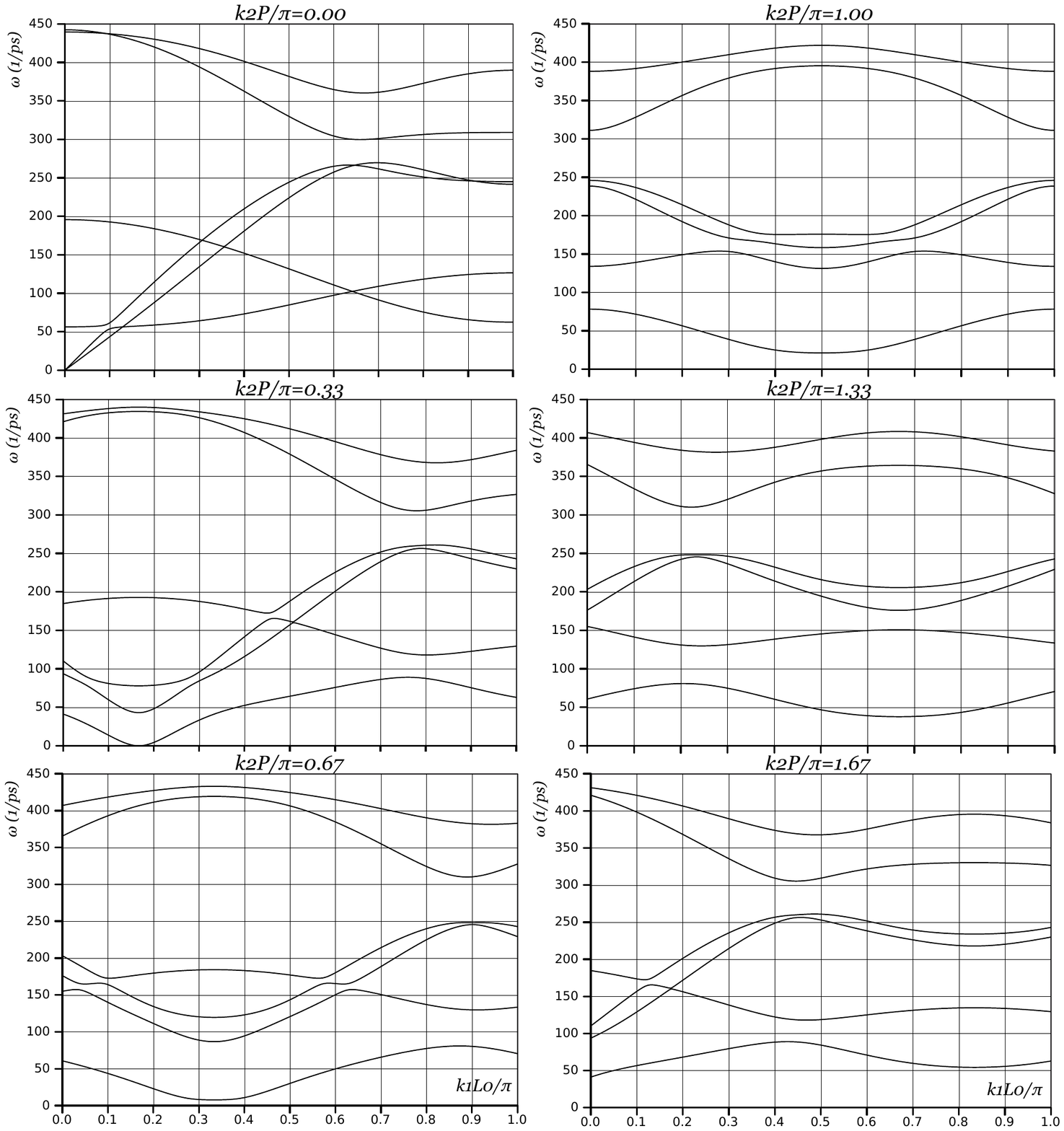}
	\caption{Dispersion curves of a $(6,6)$ carbon nanotube using Choice 4 with 2 atoms in the unit cell.  $k_1$ is the component of the wavevector in the continuous direction and normalized by $0.123\text{nm}$, and $k_2$ is the component in the discrete direction and normalized by the approximate perimeter $P$.}
	\label{fig:n6m6_2A}
\end{figure}

We compare the effect of using four different unit cells.
\begin{description}

 	\item[Choice 1:] We use a periodic unit cell with 24 atoms, the smallest number required for periodicity.  In OS terms, the group generators are $g_1=\text{identity}$ and a translation $g_2=(\bfI|0.246 \text{nm } \bfe)$.
	The phonon dispersion curves are plotted in Fig. \ref{fig:n6m6_24A}a.

	\item[Choice 2:] We use 24 atoms in the unit cell as in Choice 1, but in this case the images are related not by periodicity but by both translation (along $\bfe$) and rotation (around $\bfe$).  The generators are $g_1=\text{identity}$ and a screw $g_2=( \bfR_{\pi/3} | 0.246\text{nm } \bfe)$.
	The phonon dispersion curves are plotted in Fig. \ref{fig:n6m6_24A}b.

	\item[Choice 3:] We use 12 atoms in the unit cell, with generators closely related to Choice 2: $g_1=\text{identity}$ and a screw $g_2=( \bfR_{\pi/6} | 0.123\text{nm } \bfe)$
	The phonon dispersion curves are plotted in Fig. \ref{fig:n6m6_12A}.

	\item[Choice 4:] We make full use of the OS framework and use 2 atoms in the unit cell\footnote{OS constructed by non-commuting groups can describe carbon nanotubes with 1 atom per unit cell, but the complexity introduced by the non-commuting elements is formidable.}.
	The generators are  $g_1=(\bfR_{\pi/3}|\bfzero)$ and $g_2=(\bfR_{\pi/6}|0.123\text{nm } \bfe)$.
	In Choices 1, 2, 3, the wavevector was one-dimensional because the structure was indexed by a single index (not a triple-valued multi-index).
	In this case, the wavevector is two dimensional, but takes only $6$ discrete values in the direction corresponding to the rotation.
	This is because once we raise the rotation to the $6$-th power, we start over; conceptually, this is similar to a finite ring of atoms that has a finite set of normal modes.
	The phonon dispersion curves are plotted in Fig. \ref{fig:n6m6_2A}.

\end{description}

Comparing the plots in Fig. \ref{fig:n6m6_24A} shows, as expected, that there is a mapping between the plots obtained from Choices 1 and 2.
Any eigenvalue in one is also present in the other, though typically at a different wavevector.
Further, comparing Figs. \ref{fig:n6m6_24A}b and \ref{fig:n6m6_12A}, shows that if we unfold the curves of Fig \ref{fig:n6m6_24A}b we will recover Fig. \ref{fig:n6m6_12A}.
Similarly, Fig. \ref{fig:n6m6_2A} contains all the information, but in a much simpler description.

\begin{figure}[htbp]
	\centering
	\includegraphics[width=170mm]{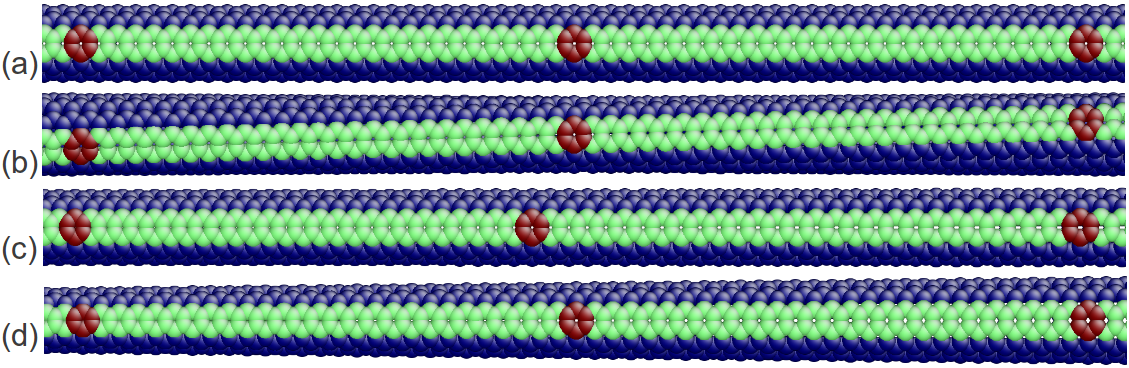}
	\caption{Phonon modes at $k_1 \approx 0$ and $k_2=0$ using Choice 4.  (a) The undeformed reference nanotube.  The colors of the atoms are only to enable visualization of the deformation.  (b) The lowest branch corresponding to twisting.  (c) The next-to-lowest branch corresponding to axial elongation.  (d) The third-from-lowest branch corresponding to a change in radius.  In each of these deformations, the cross-section remains circular.}
	\label{fig:n6m6Modes}
\end{figure}

The OS description with 2 atoms per unit cell provides a useful perspective to examine the deformations.
First, consider the case when the component of the wavevector in the discrete direction is $0$.
Each unit cell in the cross-section has the same displacement (in Objective Space) in the direction that corresponds to the discrete component of the wavevectors.
Roughly, this corresponds to ``cross-sections'' that retain their ``shape'' and remain circular.
The lowest three modes corresponding to $k_2=0$ and $k_1 \approx 0$ are plotted in Fig. \ref{fig:n6m6Modes}.

\begin{figure}[htbp]
	\centering
	\subfigure{\footnotesize (a)} \includegraphics[width=13cm, clip]{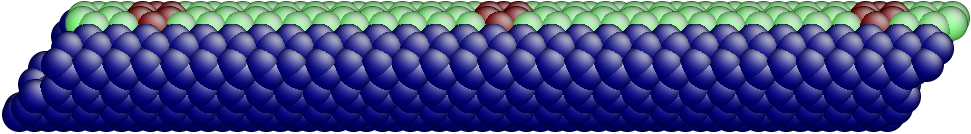} \\
	\subfigure{\footnotesize (b)} \includegraphics[width=13cm, clip]{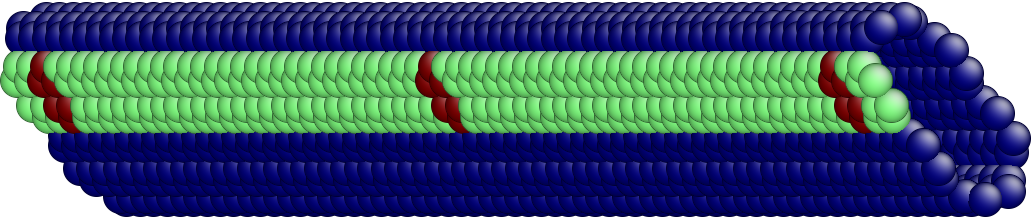} \\
	\subfigure{\footnotesize (c)} \includegraphics[width=13cm, clip]{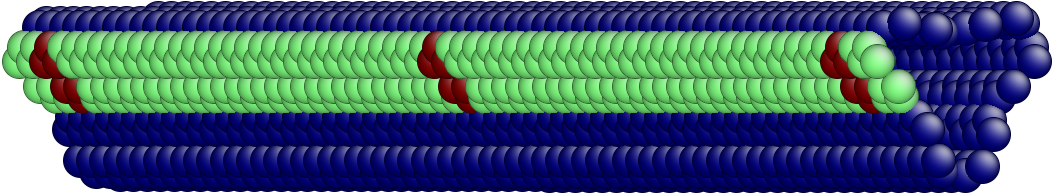} \\
	\subfigure{\footnotesize (d)} \includegraphics[width=13cm, clip]{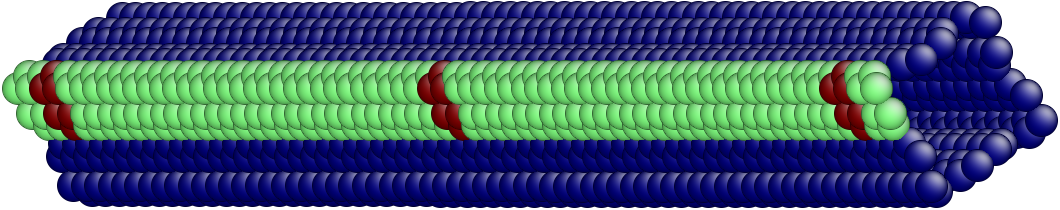} \\
	\subfigure{\footnotesize (e)} \includegraphics[width=13cm, clip]{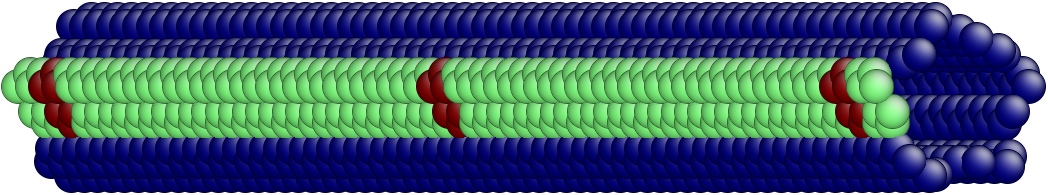}
	\caption{Some long-wavelength modes at finite $k_1$ and $k_2$ using Choice 4 (Fig. \ref{fig:n6m6_2A}).  The symmetry of the cross-section corresponds to the value of the discrete component of the wavevector.  (a) Mode from second branch at $k_1 L_0 / \pi= 1/6$ and $k_2 P / \pi= 1 /3$, similar to warping,  (b) Mode from first branch at $L_0 k_1 / \pi = 1/3$ and $P k_2 / \pi =2/3$, (c) Mode from first branch at $ k_1  L_0 / \pi= 1/2$  and $P k_2 / \pi= 1$, (d) Mode from first branch at $L_0 k_1 / \pi= 2/3$ and $P k_2 / \pi= 4/3$, (e) Mode from first branch at $L_0 k_1 / \pi= 5/6$ and $P k_2 / \pi=5/3$.}
	\label{fig:n6m6Modes2}
\end{figure}

Next, consider the case when $k_2$, the component of the wavevector in the discrete direction, is non-zero.
In this case, the cross-sections no longer remain circular.
Fig. \ref{fig:n6m6Modes2} shows examples of these modes.
Notice the relation between $k_2$ and the symmetry of the cross-section.

\begin{figure}[htnp]
	\centering
	\includegraphics[height=15mm,angle=90]{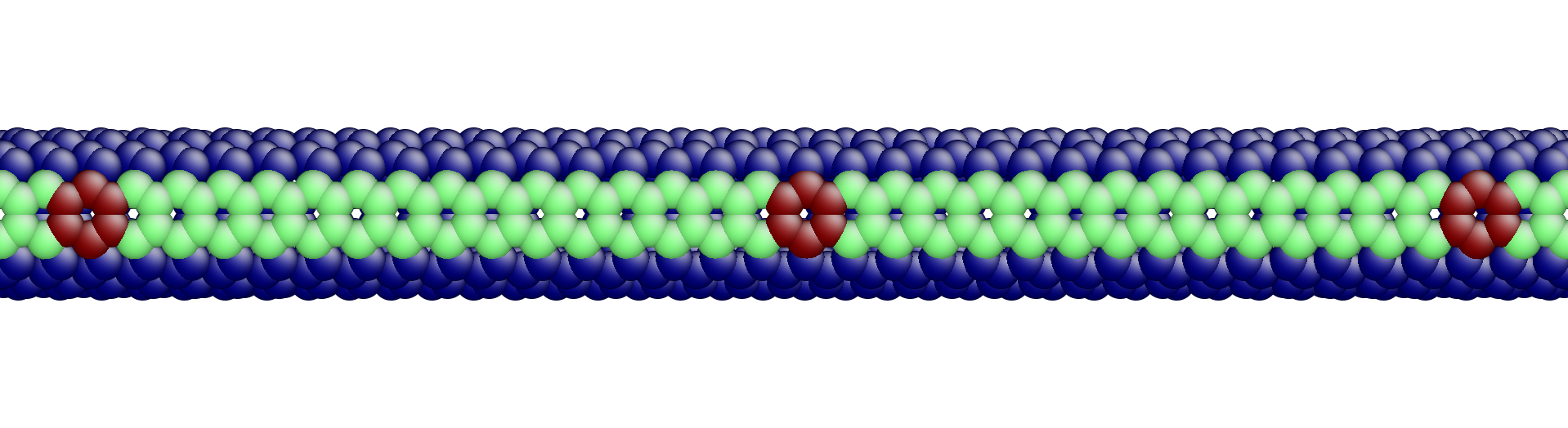}
	\includegraphics[height=15mm,angle=90]{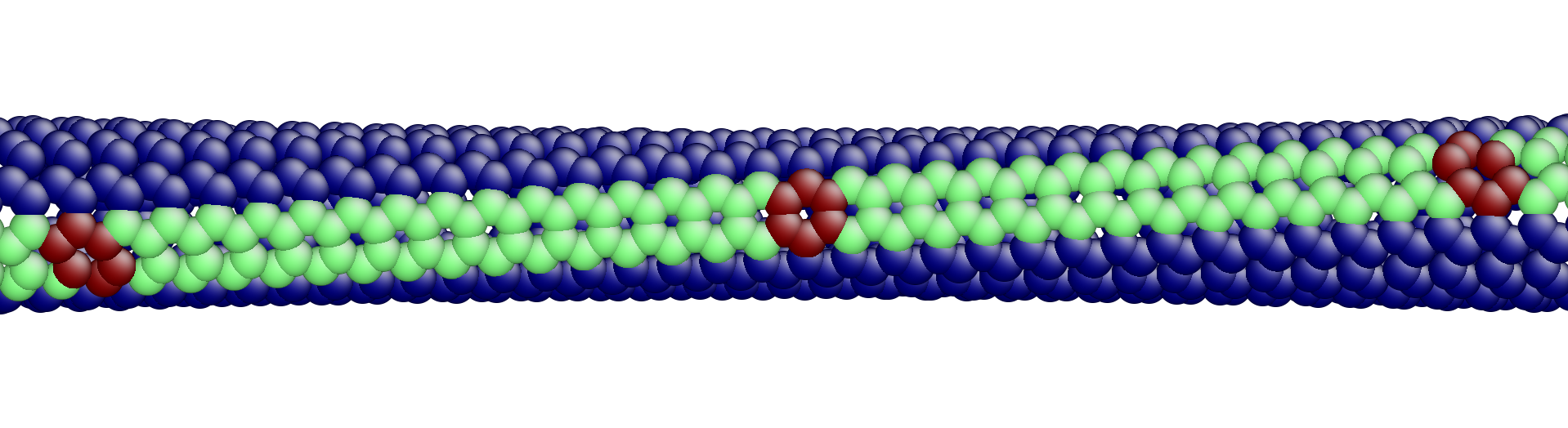}
	\includegraphics[height=15mm,angle=90]{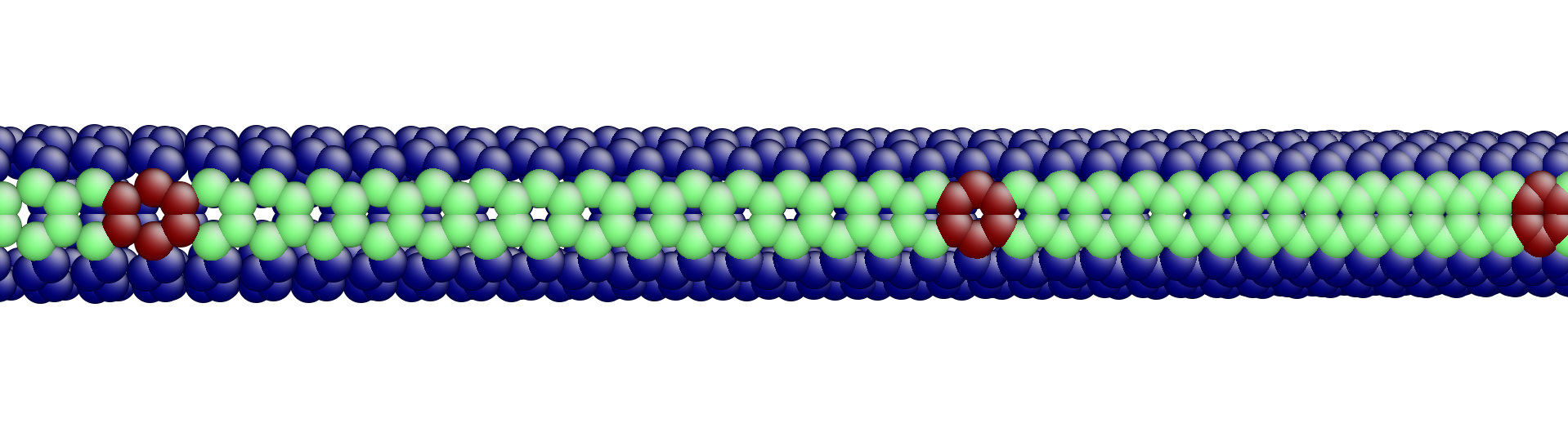}
	\includegraphics[height=15mm,angle=90]{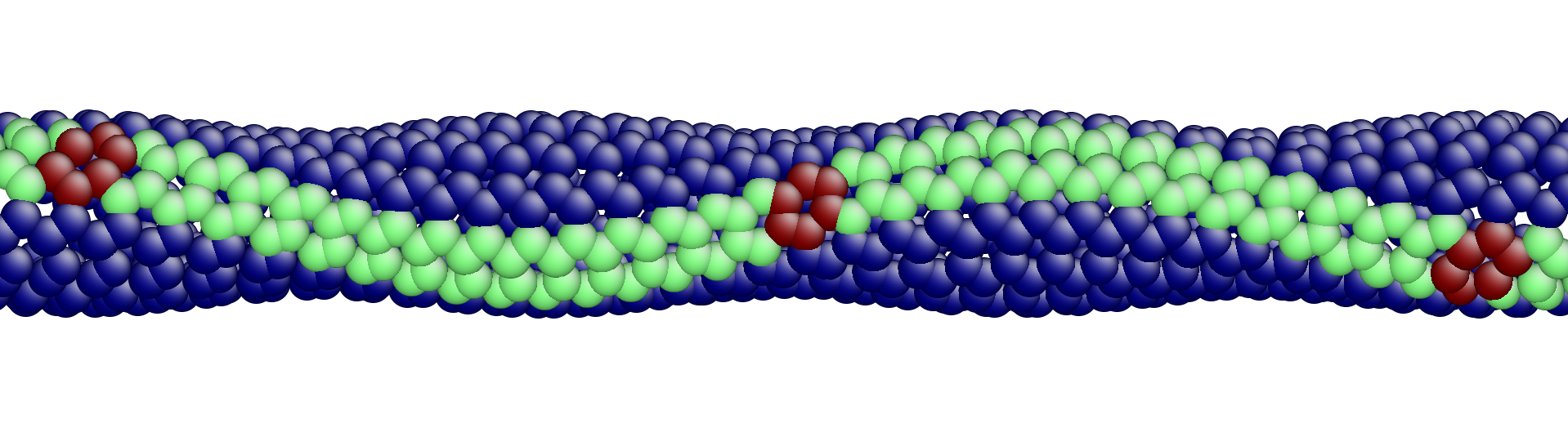}
	\includegraphics[height=15mm,angle=90]{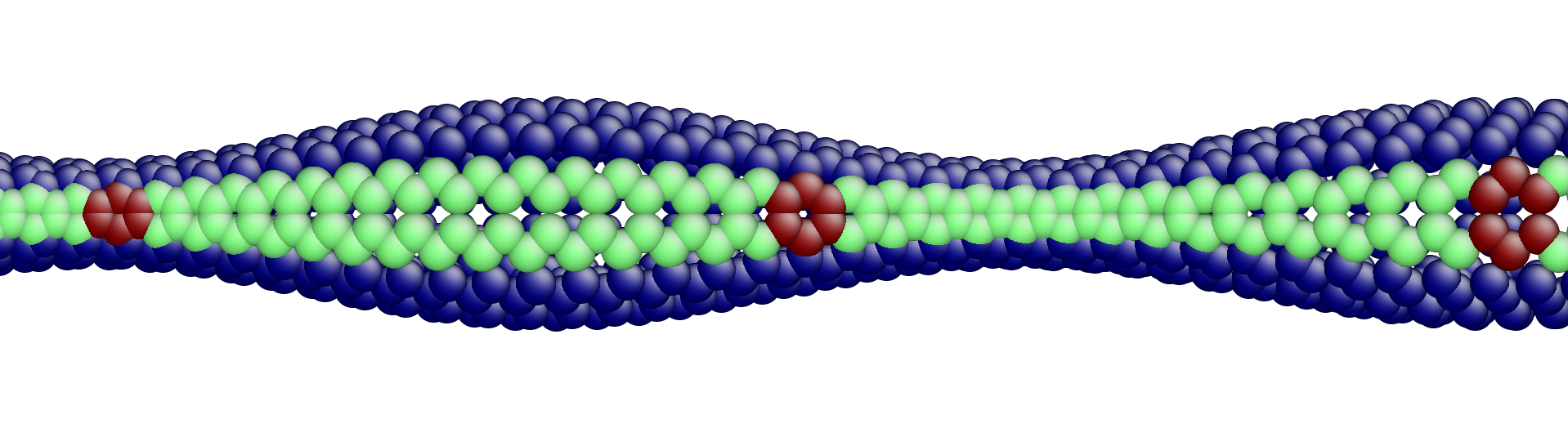}
	\includegraphics[height=15mm,angle=90]{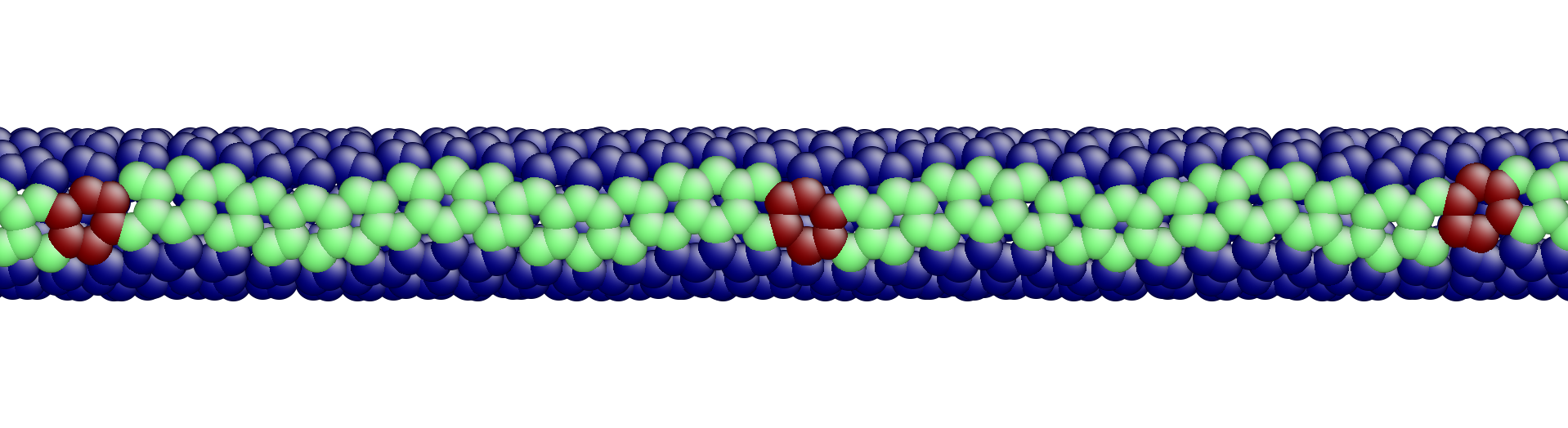}
	\includegraphics[height=15mm,angle=90]{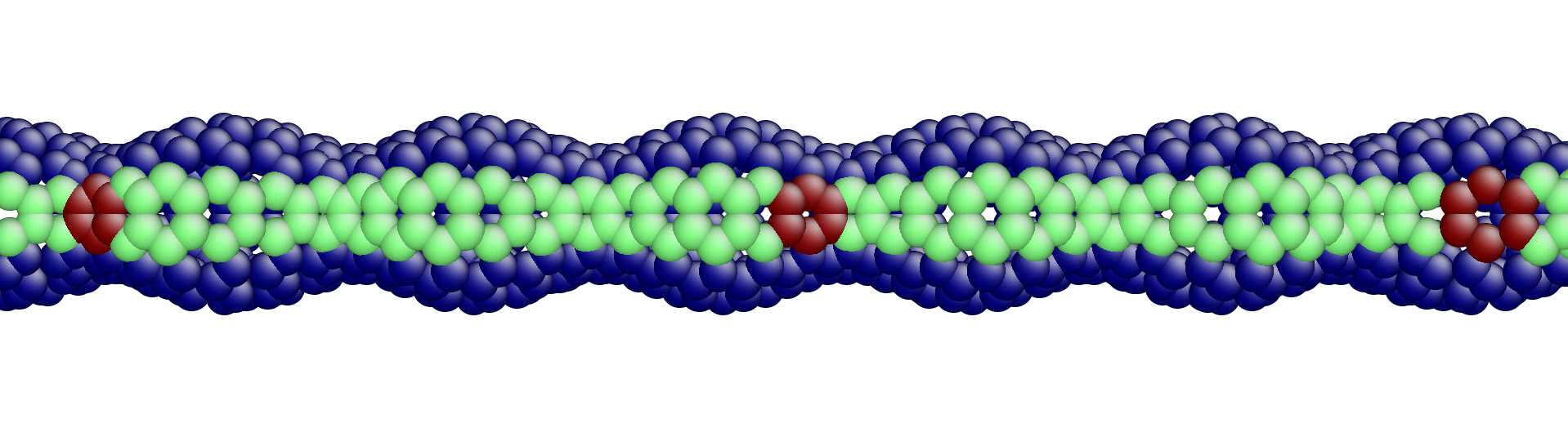}
	\includegraphics[height=15mm,angle=90]{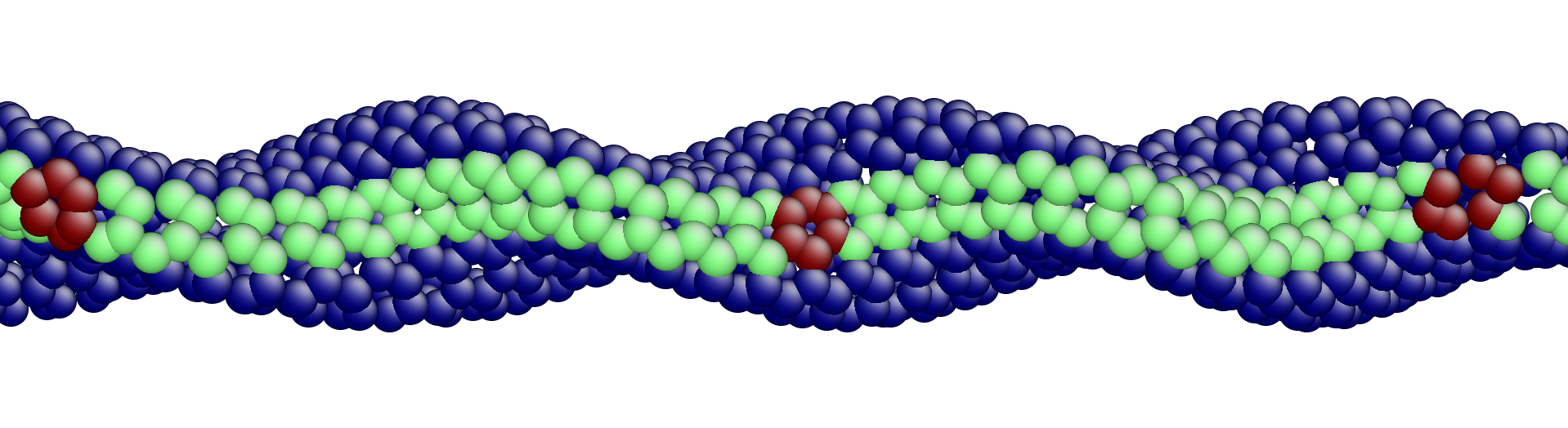}
	\includegraphics[height=15mm,angle=90]{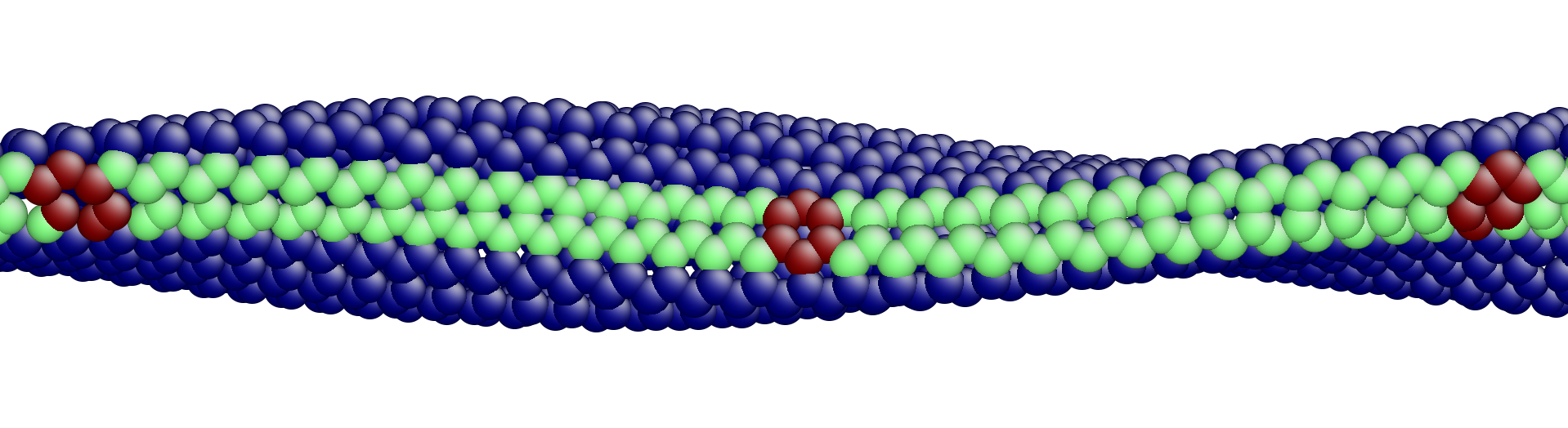}
	\caption{\small An assortment of phonon modes at finite wavevectors.  First figure is the reference state.  The colors of the atoms have no significance and are only to enable easy visualization.}
	\label{fig:nanotube-phonons-modes}
\end{figure}

Fig. \ref{fig:nanotube-phonons-modes} shows an assortment of generic phonon modes at finite wavevectors.

\subsection{Density of States (DoS) of a $(6,6)$ carbon nanotube}

\begin{figure}[H]
	\centering
	\includegraphics[trim=10mm 143mm 10mm 29mm, height=9cm, clip]{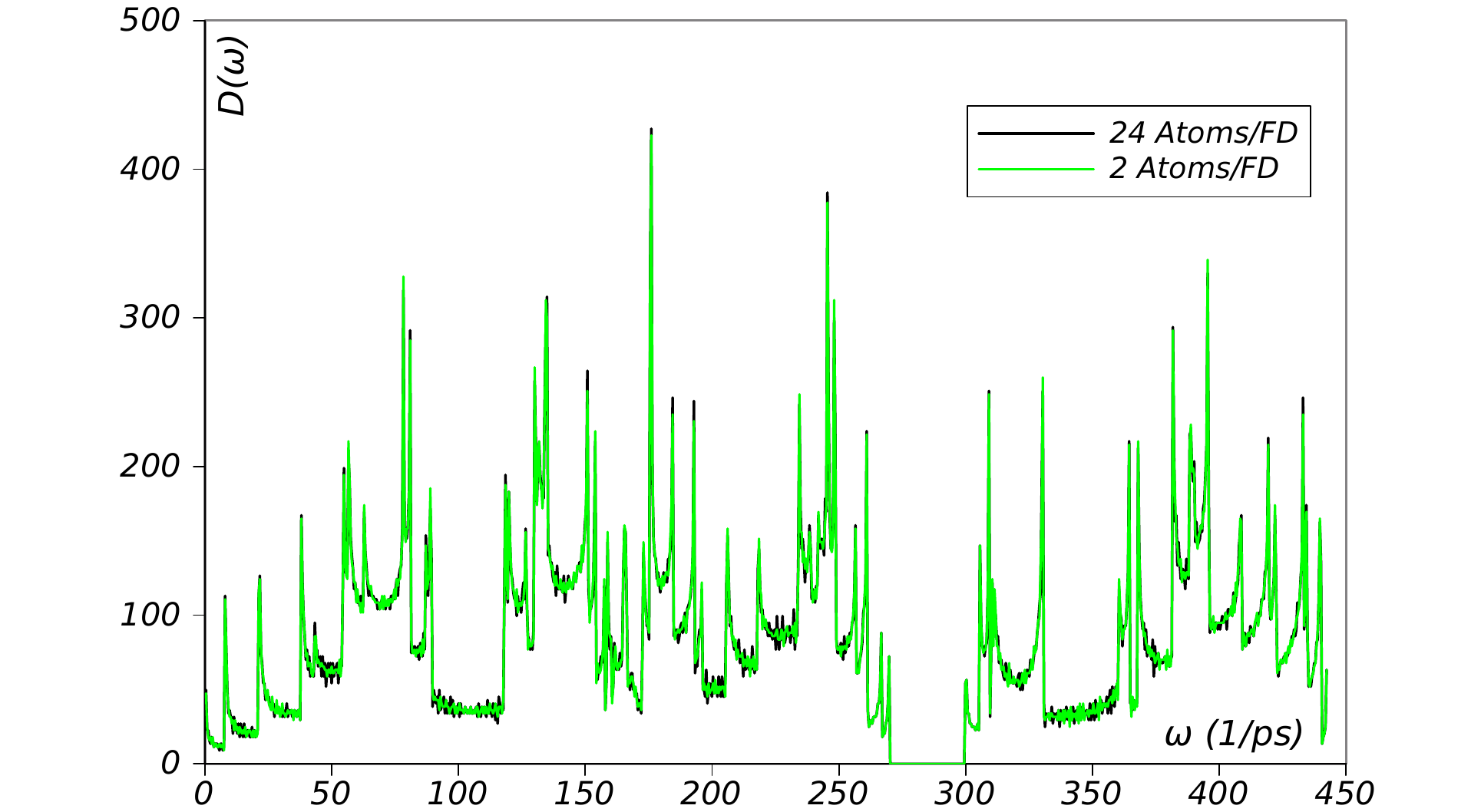}
	\caption{Density of states of a $(6,6)$ nanotube.}
	\label{fig:DOS}
\end{figure}

The density of states (DoS) of a system is an important thermodynamic quantity.
It describes the number of modes or states available per unit energy (or frequency) at each energy level.
The DoS can be calculated by making a histogram of the phonons frequencies of the system.
Fig. \ref{fig:DOS} shows the DoS of a $(6,6)$ carbon nanotube constructed using Choices 1 and 4 for the unit cell.
As expected, these curves are identical but the OS approach requires much less computational effort.

\section{Long Wavelength and Rigid Body Modes for Carbon Nanotubes} \label{long-waves}

In a 3D periodic crystal lattice, the lowest three eigenvalue branches tend linearly to $0$ as $\bfk \rightarrow {\bf 0}$.
These {\em acoustic modes} correspond to uniform deformations with rigid body translation modes as the limit deformation.
We find unusual contrasts with the crystal case when we apply this to carbon nanotubes.
We find rigid body (zero energy) motions at both zero and finite wave vectors; in addition, we find that the long-wavelength deformation corresponding to uniform radial expansion costs finite energy in real-space even in the limit of $\bfk \rightarrow {\bf 0}$, thereby giving only two eigenvalue branches that tend to $0$.
The essential explanation for these observations is that long-wavelength is now defined with respect to objective space and not real space, whereas rigid-body modes are  posed in real space for physical reasons.

We first outline this issue using as an example the choices of unit cell from Section \ref{Example}.

In Choice 1, four branches start from the origin, Fig. \ref{fig:n6m6_24A}a.
These correspond to (i) axial stretch / translation with uniform motion along $\bfe$, (ii) twist / rotation with uniformly tangential motion, and (iii) bending / translation in the plane normal to $\bfe$.
The bending mode is characterized by two degenerate branches with zero slope.
The degeneracy is due the subspace of translations in the plane being two-dimensional \cite{aghaei-dayal-elliott-inprep}.

In Choice 2, Fig. \ref{fig:n6m6_24A}b, only two branches start from the origin.
One corresponds to axial stretch / elongation, and the other one corresponds to twist / rotation.
There is a branch that has zero frequency at $k L_0 / \pi = 1/3$.
This corresponds to the rigid translation modes in the plane normal to $\bfe$, as we examine below.

Choice 3, Fig. \ref{fig:n6m6_12A}, is very similar to Choice 3, except that the branch with zero frequency at finite wavevector now goes to zero at $k L_0 / \pi = 1/6$.
We recall that that $L_0$ differs in Choices 2 and 3 precisely by a factor of 2, therefore this shift is simply because of unfolding the band diagram.

Choice 4, Fig. \ref{fig:n6m6_2A}, shows the rigid translation modes at $k_2 P / \pi = 1/3$.
In addition, for $k_2 P / \pi = 0$, we see only two branches that go to $0$ as the wavevector tends to zero; the lowest non-zero branch corresponds to uniform radial motion of the atoms.
Because every unit cell has precisely the same deformation (in Objective space), this appears at $\bfk \rightarrow {\bf 0}$.
In addition, because the atoms within the unit cell do not move relative to each other, this corresponds to the acoustic modes that are zero energy at zero wavevector in crystals.
The three higher branches at zero wavevector have the atoms in the unit cell moving with respect to each other, i.e. optic modes, and these are expected to have finite energy at zero wavevector.



\subsection{Long Wavelength Modes in Objective Space}

A long wavelength mode in Objective Space corresponds to $\bfk \rightarrow {\bf 0}$.
However, uniform deformations or their limiting rigid body translation / rotation modes, do not have as close a correspondence with long wavelengths as in crystal, because of the intermediate transformation to Objective space.
Consider a deformation induced by a normal mode, with the displacement in real space of atom $(\bfi, j)$ denoted by $\bfu_{(\bfi,j)}$.
Denote the corresponding displacement in Objective space by $\bfv_{(\bfi,j)}$.
For a normal mode with wavevector $\bfk_0$, if we set that the displacements in Objective space of corresponding atoms in every unit cell are the same, i.e. $\hat \bfv_{(\bfp,m)} = \hat \bfv_{(\bfq,m)}$ for every $(\bfp,m)$ and $(\bfq,m)$, then the DFT from Appendix \ref{DFT} gives:
\begin{equation} \label{RBMv2}
	\tilde\bfv_{(\bfj,m)} \exp \big[ -i \bfk_\bfj \cdot \bfy_\bfp \big]
	= \tilde\bfv_{(\bfj,m)} \exp \big[ -i \bfk_\bfj \cdot \bfy_\bfq \big]
	\Rightarrow
	\tilde\bfv_{(\bfj,m)}
	= \exp \big[ i \bfk_\bfj \cdot (\bfy_\bfp - \bfy_\bfq) \big] \tilde\bfv_{(\bfj,m)}
	\Rightarrow
	\bfk_0 = {\bf 0}
\end{equation}

\begin{figure}[htbp]
	\centering
	\includegraphics[width=90mm]{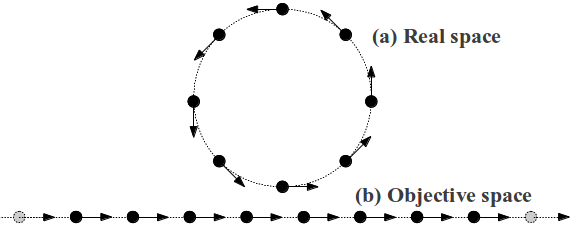}
	\caption{A long-wavelength mode that corresponds to rigid rotation and therefore zero energy: (a) A schematic projection, viewed along the axis, of atomic positions and displacements in real space.  All displacements are tangential.  (b) In Objective space, all atoms displace uniformly, i.e., long wavelength.}
	\label{fig:RBR}
\end{figure}

We consider two illustrative modes.
Fig. \ref{fig:RBR} shows schematically the position and displacements movement of atoms in real and Objective space for a rigid rotation mode.
Assume that all atoms within the unit cell translate uniformly, i.e., this is an acoustic-like mode.
It is long-wavelength in Objective space, and rigid body rotation in real space with zero energy in the limit.

\begin{figure}[htbp]
	\centering
	\includegraphics[width=90mm]{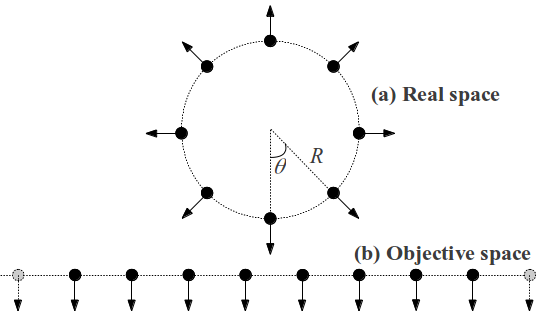}
	\caption{A long-wavelength mode that corresponds to uniform expansion rotation and therefore finite energy: (a) A schematic projection, viewed along the axis, of atomic positions and displacements in real space.  All displacements are radial.  (b) In Objective space, all atoms displace uniformly, i.e., long wavelength.}
	\label{fig:UniExpan}
\end{figure}

Fig. \ref{fig:UniExpan} shows schematically the position and displacements movement of atoms in real and Objective space for a uniform expansion mode.
Assume that all atoms within the unit cell translate uniformly, i.e., this is an acoustic-like mode.
In Objective space, this is long wavelength, but in real space this deformation costs finite energy (proportional to the square of the amplitude) even in the long-wavelength limit.

\subsection{Uniform Deformations and Rigid Body Translation in Real Space}
\label{RBT}

We now consider setting up a uniform deformation, or rather the rigid body limiting translation, in real space and then analyze the wavevector at which it appears.

As above, denote the real-space displacement of atom $(\bfp, m)$ by $\bfu_{(\bfp,m)}$ induced by a normal mode, and denote the corresponding displacement in Objective space by $\bfv_{(\bfp,m)}$.
Consider a rigid body translation mode in real space, i.e., for any two atoms $(\bfp,m)$ and $(\bfq,n)$, we have $\hat \bfu_{(\bfp,m)} = \hat \bfu_{(\bfq,n)}$.
We now find the wave vector $\bfk_\bfj$ that corresponds to this deformation.

Using (\ref{Polariz2}):
\begin{equation} \label{RBT2}
	\bfQ_{\bfp} \tilde\bfv_{(\bfj,m)} \exp \big[ -i \bfk_\bfj \cdot \bfy_\bfp \big]
	= \bfQ_{\bfq} \tilde\bfv_{(\bfj,n)} \exp \big[ -i \bfk_\bfj \cdot \bfy_\bfq \big]
	\Rightarrow
	\bfQ_{\bfp-\bfq} \tilde\bfv_{(\bfj,m)}
	= \exp \big[ i \bfk_\bfj \cdot (\bfy_\bfp - \bfy_\bfq) \big] \tilde\bfv_{(\bfj,n)}
\end{equation}
If $\bfp=\bfq$, then $\bfQ_\bfzero=\bfI$ implying that $\tilde\bfv_{(\bfj,m)}=\tilde\bfv_{(\bfj,n)}$.

Now assume $\bfr:=\bfp-\bfq \ne \bfzero$, implying $\bfy_{\bfr}=\bfy_{\bfp}-\bfy_{\bfq}$, giving the complex eigenvalue problem:
\begin{align} \label{RBT3}
	\bfQ_{\bfr} \tilde\bfv_{(\bfj,m)} = & \exp \big[ i \bfk_\bfj \cdot \bfy_\bfr \big] \tilde\bfv_{(\bfj,m)}
\end{align}
Recall that in nanotubes (Appendix \ref{SWNT-parameters}), the orthogonal part of the generators are coaxial and the axis further coincides with the nanotube axis $\bfe$.
Therefore,
\begin{equation}
	\bfQ_{\bfr\equiv(r_1,r_2)}=\bfR_{\theta_1}^{r_1} \bfR_{\theta_2}^{r_2} =\bfR_{r_1 \theta_1 + r_2 \theta_2}
\end{equation}
The eigenvalues of $\bfQ_\bfr$ are therefore $1$ and $e^{\pm i (r_1 \theta_1 + r_2 \theta_2)}$, where $\theta_1$ and $\theta_2$ are the group parameters for the nanotube (Appendix \ref{SWNT-parameters}).

There are therefore three modes corresponding to rigid body translation:
\begin{itemize}

	\item $\lambda_1=e^{i \bfk_\bfj \cdot \bfy_\bfr}=1$.
	Since this holds for all $\bfy_\bfr$,  the wavevector $\bfk_\bfj$ is zero.
	The eigenvector $\tilde\bfv_{(\bfj,m)}$ will coincide with $\bfe$,  and from (\ref{Polariz2}) it follows that all the atoms will move axially.
	This mode is rigid translation in the axial direction.
	
	\item $\lambda_2=e^{i \bfk_\bfj \cdot \bfy_\bfr}=e^{i(r_1 \theta_1 + r_2 \theta_2 )}$.
	From (\ref{XandK-1},\ref{XandK-2}), we have that $\frac{2\pi}{N_1} r_1 j_1 + \frac{2\pi}{N_2} r_2 j_2 = r_1 \theta_1 + r_2 \theta_2$ for all $r_1$ and $r_2$.
	Therefore, $\frac{2\pi j_1}{N_1}=\theta_1$ and $\frac{2\pi j_2}{N_2}=\theta_2$.
	That is, the wavevector at which this rigid body translation occurs is $k_1=\theta_1, k_2 = \theta_2$.
	The eigenvector $\tilde\bfv_{(\bfj,m)}$ is orthogonal to the first eigenvector $\bfe$.
	In addition, using $\bfQ_\bfr \tilde\bfv_{(\bfj,m)}= e^{i \theta} \tilde\bfv_{(\bfj,m)}$ into (\ref{Polariz2}), we find that the nanotube will rigidly translate in the plane with normal $\bfe$.

	\item $\lambda_3=e^{i \bfk_\bfj \cdot \bfy_\bfr}=e^{-i(r_1 \theta_1 + r_2 \theta_2 )}$.
	As with $\lambda_2$, the wavevector at which this rigid body translation occurs is $k_1=- \theta_1, k_2 = - \theta_2$.
	Since the wavevector is meaningful only up to sign, this is essentially the same.
	The eigenvector is also orthogonal to $\bfe$ and can be chosen normal to the second eigenvector.
\end{itemize}
The latter two modes above can alternately be considered as the limiting behavior of rigid rotations around axes that are perpendicular to $\bfe$.

\begin{figure}[htbp]
	\centering
	\includegraphics[width=90mm]{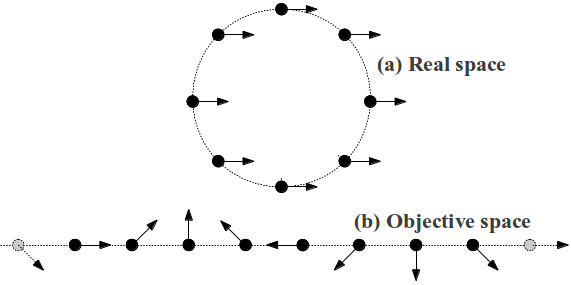}
	\caption{A rigid body translation with zero energy that corresponds to finite-wavelength: (a) A schematic projection, viewed along the axis, of atomic positions and displacements in real space.  (b) In Objective space, it is not long wavelength.}
	\label{fig:RBT}
\end{figure}

The phonon frequency of all of these modes is zero because rigid motions in real-space do not cost energy.
Fig. \ref{fig:RBT} demonstrates a schematic of a rigid body translation in real space that has finite wavelength in Objective space.
Heuristically, the Objective transformation goes to a space that ``unwraps'' the structure.

\section{Phonons and Stability}
\label{stability}

Phonon analysis provides important insights into the stability of crystals through identifying {\em soft modes}, i.e.,  non-rigid deformations that cost no  energy  \cite{dove-book}.
In addition, the phonon framework provides important insights and enables systematic identification of the appropriate larger unit cells at instabilities \cite{elliott-jmps2006a,elliott-jmps2006b}.
As discussed in \cite{elliott-jmps2006a}, phonon analysis does {\em not} provide information about stability with respect to certain deformation modes; in particular, phonon stability does not test against {\em non-rank-one modes}.
The analogy in linear continuum elasticity is that strong ellipticity tests only that waves speeds are real in all directions and all polarizations.
In terms of the stiffness tensor, this does not test positive-definiteness of the stiffness against all tensors in the 6-dimensional strain space; rather it tests only against the subspace of strains that are symmetrized rank-one tensors.
While phonons do test if solids are stable to uniform uniaxial extensions in every direction, they do not test if they are stable to superpositions of these, such as biaxial and triaxial stretch.
Because the Fourier transform does not exist for the limit deformation, superposing modes and taking the limit is not equivalent to taking the limit and then superposing.

We also note that phonons test only the material stability but not against structural instabilities such as buckling \cite{muller-triantafyllidis-geymonat}.
Structural instabilities are typically very sensitive to boundary conditions, e.g. the elementary Euler buckling loads.
Testing the linear stability of an atomic structure against structural modes requires, in general, the brute-force solution of the full eigenproblem with a very large number of degrees of freedom.

In this section, we discuss two findings relevant to the role of phonons and stability.
First, we discuss why there do not exist the analog of non-rank one modes in carbon nanotubes.
That is, assuming that all phonon branches are positive, and in addition those branches that tend to $0$ at long wavelength have positive slope (in the case of twisting and axial extension) or have positive second derivative (in the case of bending), then the nanotube is stable under any combination of these.
In other words, a positive torsional modulus, extensional modulus, and bending modulus, {\em do} imply, unlike crystals, that they are stable under any combination of torsion-extension-bending.
The second finding that we discuss is a numerical study of torsional buckling using two unit cells, one with $456$ atoms and another with $24$ atoms.
As the former choice has much more freedom in deforming, we see torsional instabilities.
With the latter choice, we find a signature of this instability in terms of zero phonon frequencies; in addition the eigenmode corresponding to the zero frequency predicts the nature of the instability.
This also displays an important calculation that is enabled by the OS framework: torsion is simply not possible with periodic boundary conditions.


\subsection{Stability Under a Combination of Long-wavelength Modes}

The fact that there do not exist analogs of non-rank one modes in carbon nanotubes is made clear by the use of the OS framework.
The OS description shows that nanotubes are one-dimensional in an essential way, in particular, the wavevector has only one continuous component.
This enables a simple calculation to show that any superposition of twisting, extension and bending must be stable, if they are each individually stable.

First, we make a note about bending of nanotubes.
As shown in \cite{dayal-elliott-james-formulas}, a nanotube that bends does not remain an OS if the the group description has no generators that are translations.
The difficulty with this situation is that atomic environments are no longer related, and in particular a theorem by James \cite{james-jmps2006} that equilibrium of single unit cell implies equilibrium of the OS is not valid.
Therefore, in such a group description, it is not possible to define a bending modulus since this requires microscopic equilibrium to be meaningful.
Alternately, any nanotube, even if chiral, can be described by a translational unit cell, though this cell may be very large.
In this description that includes a translational generator, bending is well-defined.
Essentially, it corresponds to the non-identical environments of atoms being replaced by a large unit cell in which atoms relax in possibly non-uniform ways.
However, an important feature of this OS description is that bending now occurs at $\bfk \rightarrow {\bf 0}$.
The net result is that either the bending modulus cannot be defined, or if it can be defined then bending occurs at $\bfk \rightarrow {\bf 0}$.
This is important for our calculation below.

In a 2D Bravais lattice, phonon stability tests deformations of the form
\[
	\lim_{k_1 \rightarrow 0,  k_2 = const.} \bfA_1(\bfk) e^{i k_1 x_1}, \quad \text{and } \lim_{k_2 \rightarrow 0, k_1 = const.} \bfA_2(\bfk) e^{i k_2 x_2}
\]
Here $\bfA_1$ and $\bfA_2$ are arbitrary vectors; because of linearity, we can decompose them to correspond to polarizations of the appropriate normal modes that propagate in the same direction.
Therefore, e.g., we are assured of the stability of any superposition of homogeneous shear and extension {\em only} when they are the limit of phonons that propagate in the same direction, if the component phonons are themselves stable.
However, phonon stability cannot say anything about modes that involve deformations that are superpositions of phonons that propagate in different directions, i.e., a deformation of the form
\[
 	\lim_{k_1\rightarrow 0, k_2 \rightarrow 0, k_1/k_2 = const.} \bfA_1(\bfk) e^{i k_1 x_1} + \bfA_2(\bfk) e^{i k_2 x_2}
\]
For example, uniaxial stretch in each coordinate direction can be tested by the individual limits, while biaxial deformation requires the composite limit that cannot be achieved by superposing the individual limits.

In nanotubes, we only have a single continuous component of the wavevector.
Therefore, all limits are with respect to only that component.
If deformations of the form $\displaystyle \lim_{k \rightarrow 0} \bfA_j (k) e^{i k y}$ are stable, where $\bfA_j$ corresponds to axial stretch, twist, or bending, then it follows that deformations of the form $\displaystyle \lim_{k \rightarrow 0} \sum_j \bfA_j (k) e^{i k y}$ are also stable simply by superposition.
Physically, if we have positive bending stiffness, positive torsional stiffness, and positive extensional stiffness, the nanotube is stable to any combination of bending, torsion, and elongation.


\subsection{Torsional Instabilities of Nanotubes}

Soft-mode techniques to detect instabilities at the crystal-level have a long history in mechanics, as far back as \cite{hill1977principles}.
Recently, they have been combined with bifurcation techniques to understand structural transformations in shape-memory alloys \cite{elliott-jmps2006a}.
They have also proved useful in understanding defect nucleation and propagation at the atomic scale, e.g. \cite{miller-rodney-jmps2008, lu-dayal-philmag2011, dayal-bhattacharya-jmps2006}.

We numerically study the torsional instability of a $(6,6)$ carbon nanotube using both phonons and (zero temperature) atomistics.
Phonons in principle test the stability of a large system efficiently, while atomistics requires us to use large unit cells if we are to capture complex instabilities.
We find that phonon stability provides an accurate indicator of the onset of the instability as well as the initial post-instability deformation.

\begin{figure}[htbp]
	\centering
	\subfigure{\footnotesize }\includegraphics[width=13cm]{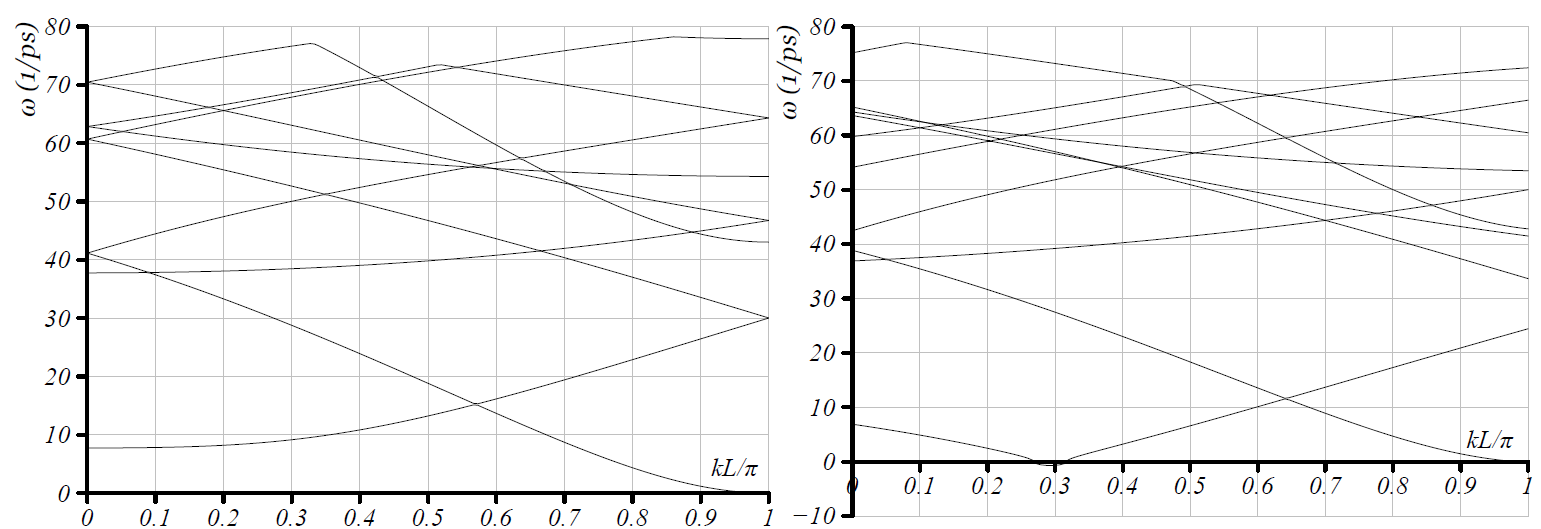}
	\subfigure{\footnotesize }\includegraphics[width=10cm, clip]{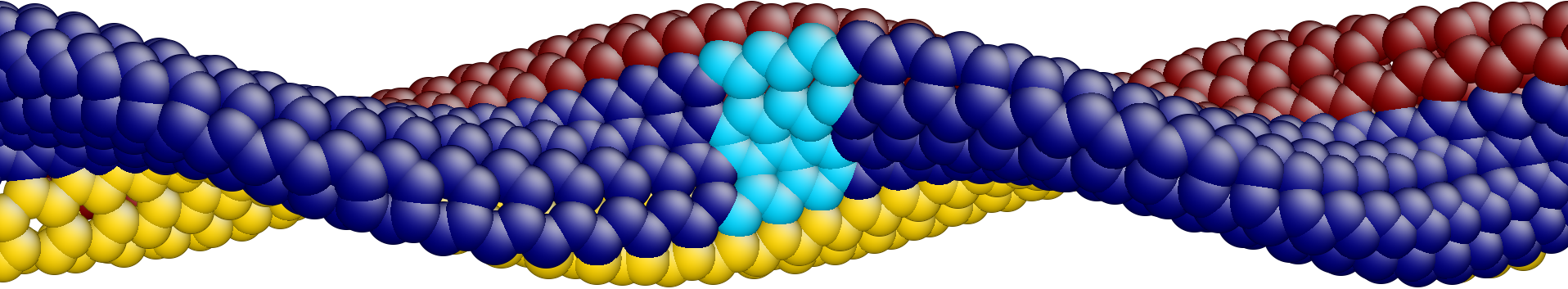}
	\subfigure{\footnotesize }\includegraphics[width=1.5cm, clip]{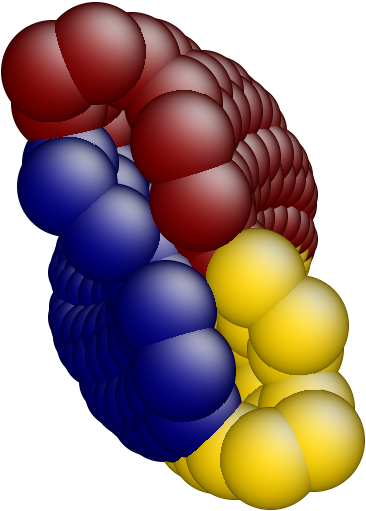}
	\caption{Phonon stability analysis. Top left: ten lowest modes for the untwisted nanotube. Top right: ten lowest modes just after the lowest eigenvalues becomes negative.  The value of the discrete wavevector is $2/3 \pi$.  The twist angle at this state is about $5^\circ/$nm.  The mode going to $0$ in both plots corresponds to long-wavelength bending.  Bottom: the eigenmode corresponding to the zero eigenvalue.  This shows the deformation predicted by phonon analysis to have zero energy.  Light blue atoms denote a single unit cell.  The colors are only to aid visualization.}
	\label{fig:TorsionBuckl}
\end{figure}

We use two different unit cells, one with $24$ atoms and the other with $456$ atoms.
The smaller unit cell requires OS group generators given by $g_1=(\bfR_{2\pi/3}|\bfzero)$ and $g_2=(\bfR_{\pi}|0.75 \text{nm } \bfe)$ and is shown in Fig. \ref{fig:TorsionBuckl}.
The larger unit cell requires a single translational generator, $g_1=(\bfI|4.8\text{nm } \bfe)$.

\begin{figure}[htbp]
	\centering
	\includegraphics[width=16cm]{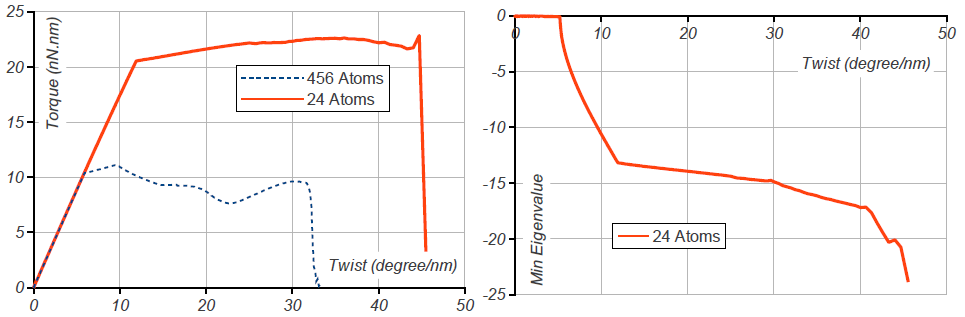}
	\caption{Left: Twisting moment vs. twist angle using atomistics. Right: Lowest eigenvalue vs. twist angle for the smaller unit cell.}
	\label{fig:TorsionTorq}
\end{figure}

For both choices, we apply a small increment of twisting moment, equilibrate, and repeat the process.
For the smaller unit cell, we additionally test the phonon stability by computing the phonon frequencies at each load step.
The twisting moment vs. twist angle and lowest eigenvalue vs. twist angle are plotted in Fig. \ref{fig:TorsionTorq}.
In the atomistic simulations, the larger unit cell buckles at much lower twist angle (about $5^\circ/$nm) compared to $12^\circ/$nm for the smaller unit cell.
However, the phonon analysis of the smaller unit cell indicates that an eigenvalue becomes negative at about $5^\circ/$nm.
This is consistent with the onset of buckling for the larger unit cell.
Additionally, the eigenmode corresponding to the negative eigenvalue matches with buckling mode of the long tube computed directly from atomistics.
The atomic deformation corresponding to the eigenmode is plotted in Fig. \ref{fig:TorsionBuckl}, along with phonon spectra before and at the point of instability.

This calculation also provides a method to test for one possible route to failure for the OS analog of the Cauchy-Born rule.
Specifically, loss of phonon stability is an indicator that the unit cell must be enlarged, i.e. affinely applied far-field boundary conditions do {\em not} give affine deformations of each unit cell \cite{Friesecke-Theil}.

\section{Energy Transport in Helical Objective Structures} \label{transport}

Motivated by the features of the computed phonon curves in nanotubes, in this section we present a simplified geometric model that aims to capture the key physics of energy transport.
The model is based on a balance between energy transport along a helical path and energy transport along an axial path.
In an ``unwrapped'' helix, the former corresponds to transport through short-range interactions, i.e. the interactions are between neighbors that are nearby in terms of the labeling index, and the latter corresponds to long-range interactions, i.e. the interactions are between distant atoms in terms of the labeling index.
Of course, in physical space, both these types of neighbors are at comparable distances, and interactions are therefore of comparable strength.

\begin{figure}[htbp]
	\centering
	\includegraphics[trim=8mm 71mm 5mm 72mm, height=9.5cm, clip]{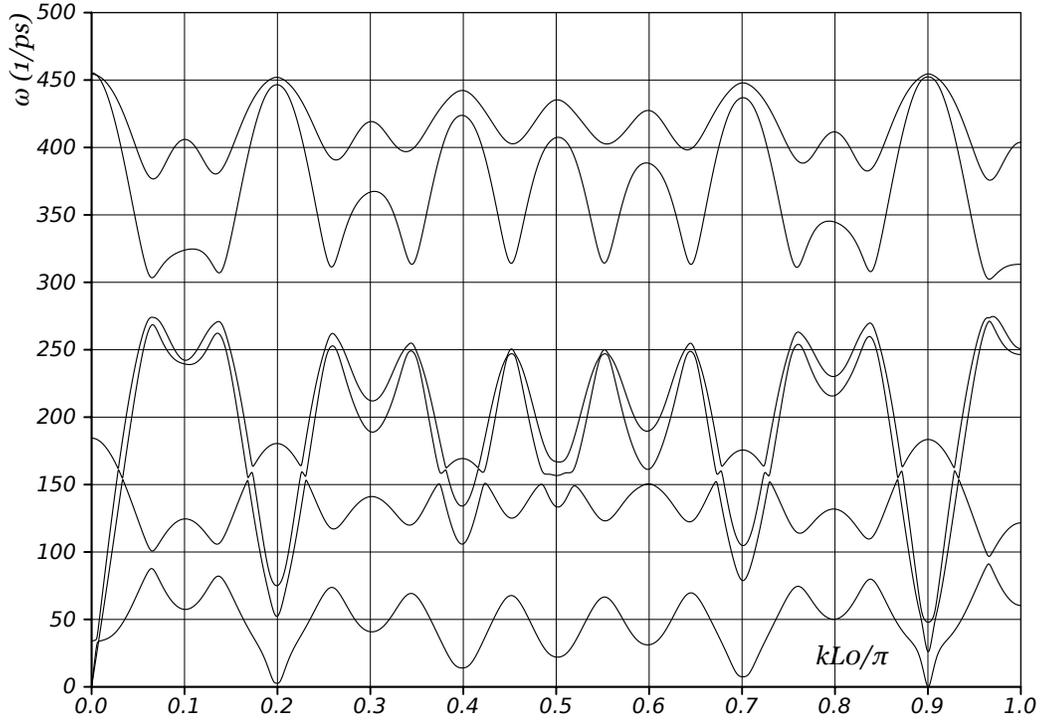}
	\caption{Dispersion curves of a $(11,9)$ carbon nanotube. FD contains 2 atoms.}
	\label{fig:n11m9}
\end{figure}

\begin{figure}[htbp]
	\centering
	\subfigure{\footnotesize (a)}\includegraphics[trim=8mm 71mm 8mm 72mm, height=9cm, clip]{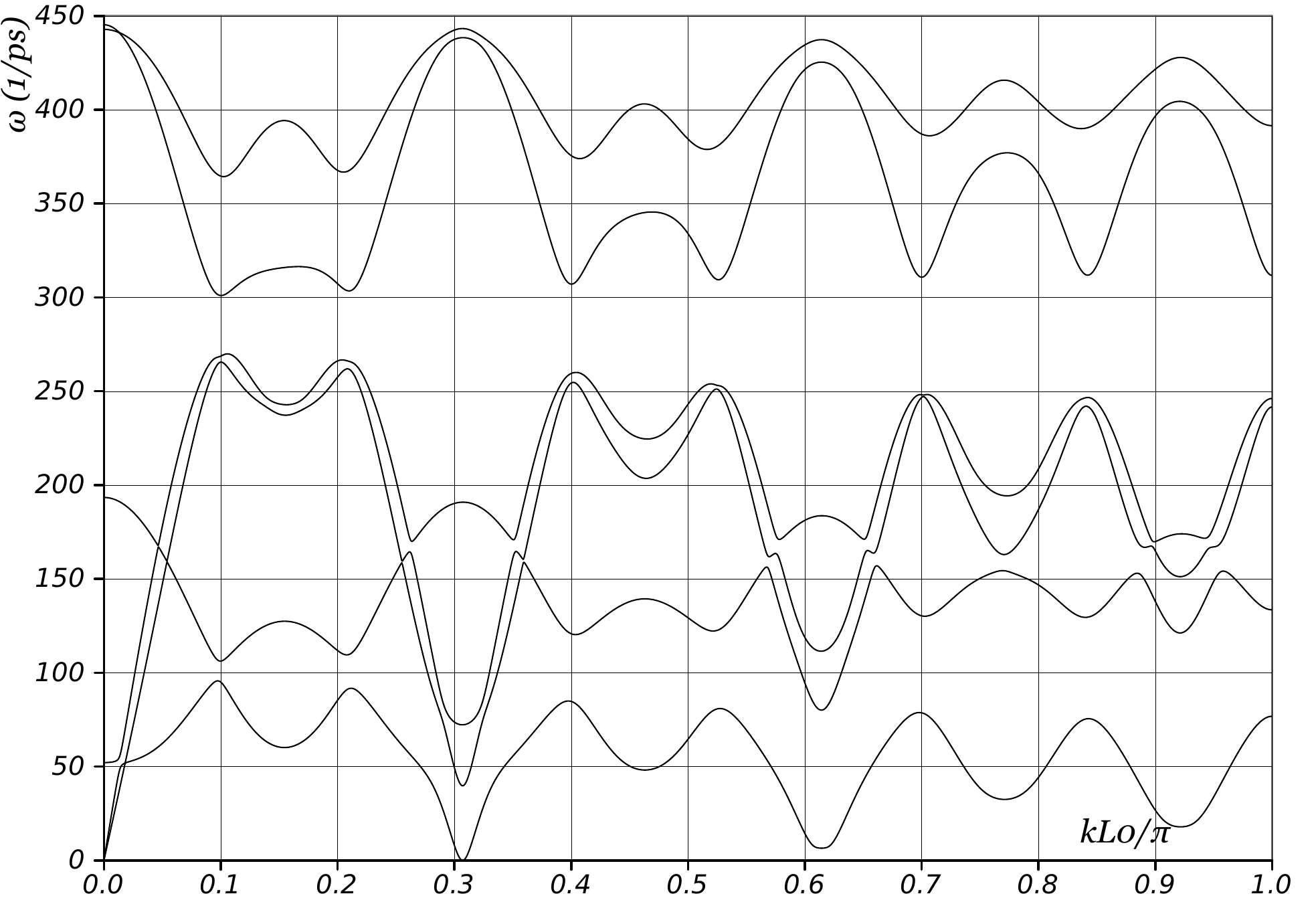}
	\subfigure{\footnotesize (b)}\includegraphics[trim=8mm 71mm 8mm 72mm, height=9cm, clip]{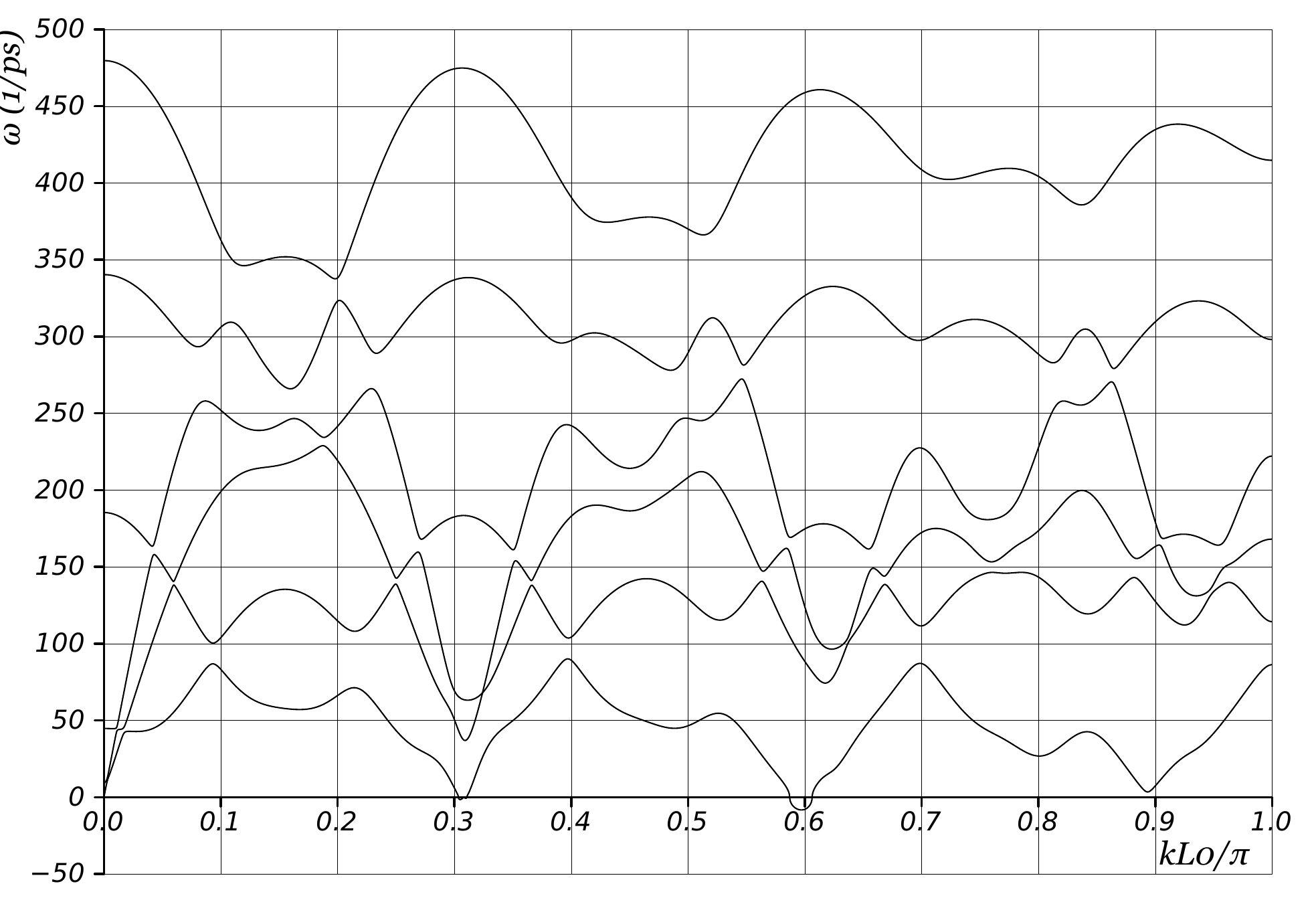}
	\caption{Dispersion curves of a $(7,6)$ carbon nanotube. FD contains 2 atoms. (a) relaxed nanotube (b) compressed and twisted nanotube.}
	\label{fig:n7m6}
\end{figure}

We begin by examining the phonon curves of nanotubes $(m,n)$ where $m$ and $n$ are relatively prime.
As noted in Appendix \ref{SWNT-parameters}, this implies that a single screw generator is sufficient to describe the nanotube with 2 atoms per unit cell.
Figs. \ref{fig:n11m9} and \ref{fig:n7m6}a  show the dispersion curves of unloaded (11,9) and (7,6) nanotubes respectively.
The rotation angles of the screw generator are $\theta_1=\frac{271\pi}{301} \approx 0.9003\pi$ and $\theta_1=\frac{39\pi}{127} \approx 0.307\pi$, respectively.
As discussed in previous sections, two branches corresponding to torsion and axial elongation start from the origin, and one branch touches the $k$ axis at precisely $\theta_1$.
We mention that if we had used the periodic description for these nanotubes, we would require at least $1204$ and $508$ atoms in the unit cell for the $(11,9)$ and $(7,6)$ nanotubes respectively.
Besides the significantly larger computational expense, it would imply that Figs. \ref{fig:n11m9} and \ref{fig:n7m6}a contain $3612$ and $1524$ curves respectively!
Physical interpretation would be impossible.

The key features of interest here are the ``wiggles'' in the phonon curves in Figs. \ref{fig:n11m9} and \ref{fig:n7m6}a.
There exist certain distinguished wavevectors at which the group velocity (i.e. slope of the dispersion curve) becomes zero in all branches.
These wavevectors are primarily selected by geometry: Fig. \ref{fig:n7m6} compares the curves for a $(7,6)$ nanotube both with no load as well as with compressive axial force and nonzero twisting moment.
In addition, the phonon curves depend on the specific interatomic potential, but we have found that the distinguished wavevectors have a very weak dependence.
Similar wiggles, though not as prominent are also visible in Fig. \ref{fig:n6m6_2A} for a $(6,6)$ nanotube.
Since the group velocity gives the speed of energy transport, there is no energy transport at these distinguished wavevectors.
These observations motivate a geometric model for the energy transport that neglects much of the complexity of the interatomic potential.

\subsection{A Simplified One-Dimensional Nonlocal Model for Helical Objective Structures}

\begin{figure}[htbp]
	\centering
	\includegraphics[width=60mm]{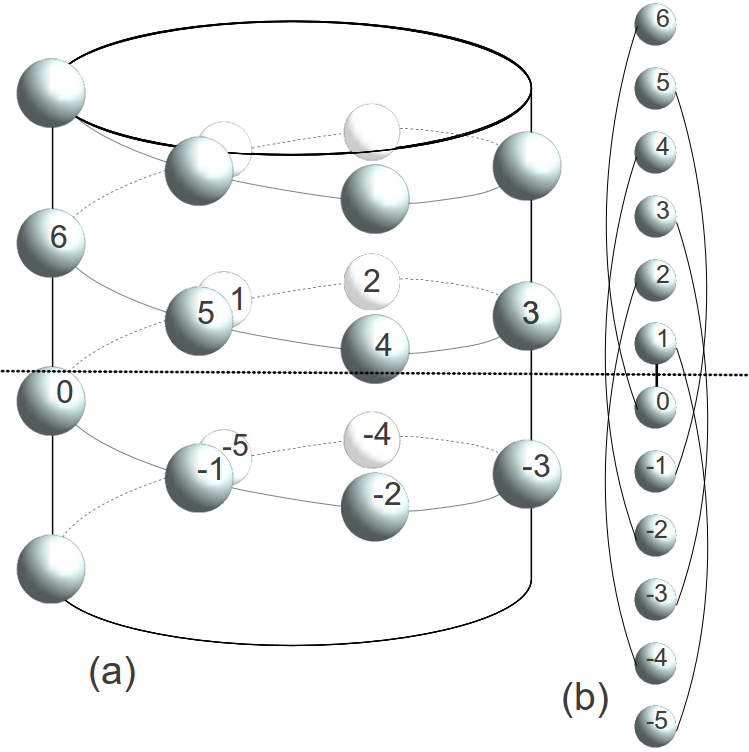}
	\caption{A schematic of the of the geometric model for energy transport  (a) In real space (b) In Objective space.  The energy flow to be analyzed takes place across the bold dashed line, i.e., how much energy do subunits $0, -1, -2, \ldots$ transfer to the subunits $1, 2, 3, \ldots$.  The subunits can correspond to individual atoms or sets of atoms.}
	\label{fig:energy-flux}
\end{figure}

The key idea is that when a helix is plotted in a space that uses the path length along the helix as the coordinate, there are short-range interactions that are due to neighbors along the helix in real space, and there are long-range interactions due to interactions between neighbors that lie above on the next loop in real space.
Fig. \ref{fig:energy-flux} shows a schematic of this geometric picture using a specific example.
The goal is to write down the expression for energy transfer to the atoms with positive labels from the atoms with non-positive labels.
Roughly, we want the energy flux crossing the surface represented by the dashed line.
In real space, the roughly equal-strength bonds that cross the dividing surface are between atom pairs $(0,1), (0,6), (-1, 5), (-2, 4), (-3,3), (-4,2),(-5,1)$.
In objective space, only the first of these bonds is ``local'' while the others are all ``non-local''.

The picture above for a generic nanotube with 2 generators is not essentially different.
The Objective space picture is a set of parallel atomic chains, with infinite length in the direction corresponding to the powers of the screw generator, but a finite number of parallel chains with the number of of chains corresponding to the powers of the rotation generator.
This can be considered as simply a single linear chain with an expanded unit cell.

For an OS $\Omega$, we write down the total energy flux $\psi$ from a subbody $\Omega^+$ to $\Omega^- := \Omega \setminus \Omega^+$ over a time interval $T$:
\begin{equation} \label{EFlux1}
	\psi = \int_t^{t+T} \sum_{(\bfp,m) \in \Omega^+} \sum_{(\bfq,n) \in \Omega^-} \dot\bfu_{(\bfp,m)} \cdot \bff_{(\bfp,m)(\bfq,n)} dt
\end{equation}
The superposed $\dot{\square}$ represents the time derivative.
The term $\bff_{(\bfp,m)(\bfq,n)}$ is the force between the atoms $(\bfp,m)$ and $(\bfq,n)$; while this is not always a uniquely-defined quantity in multibody potentials, in a linearized system this is simply $\bff_{(\bfp,m)(\bfq,n)} := \bfH_{(\bfp,m)(\bfq,n)} \left( \bfu_{(\bfp,m)} - \bfu_{(\bfq,n)} \right)$.

We now compute the energy flux for a single phonon mode $\bfu_{(\bfp,m)}=\bfQ_\bfp \hat\bfu_m \cos (\bfk \cdot \bfy_\bfp - \omega t + \vartheta)$ where $\vartheta$ is a phase that eventually gets integrated out and disappears.
We set the averaging interval to a single cycle, i.e., $T = 2\pi / \omega$.
The energy flux is therefore
\begin{equation} \label{EFlux3}
	\psi = \omega \int_t^{t+T} \sum_{(\bfp,m) \in \Omega^+} \sum_{(\bfq,n) \in \Omega^-}
	\left[ \begin{array}{c} \bfQ_\bfp \hat\bfu_m \sin (\bfk \cdot \bfy_\bfp - \omega t + \vartheta) \cdot \bfH_{(\bfp,m)(\bfq,n)} \cdot \\
	\left( \bfQ_\bfp \hat\bfu_m \cos (\bfk \cdot \bfy_\bfp - \omega t + \vartheta) - \bfQ_\bfq \hat\bfu_n \cos (\bfk \cdot \bfy_\bfq - \omega t + \vartheta) \right)
	\end{array} \right] dt
\end{equation}
Using (\ref{hessian1}), we have that
\begin{align} \label{EFlux4}
	\psi & = \frac{\omega}{2} \int_t^{t+T} \sum_{(\bfp,m) \in \Omega^+} \sum_{(\bfq,n) \in \Omega^-}  \hat\bfu_m \cdot \bfH_{(\bfzero,m)(\bfq-\bfp,n)} \hat\bfu_m \sin \big( 2\bfk \cdot \bfy_\bfp - 2\omega t + 2\vartheta \big)  dt  \notag   \\
	&  - \frac{\omega}{2} \int_t^{t+T} \sum_{(\bfp,m) \in \Omega^+} \sum_{(\bfq,n) \in \Omega^-}  \hat\bfu_m \cdot \bfH_{(\bfzero,m)(\bfq-\bfp,n)} \bfQ_{\bfq-\bfp} \hat\bfu_n  \Big[ \sin \big( \bfk \cdot (\bfy_\bfp+\bfy_\bfq) - 2\omega t + 2\vartheta \big)  \notag \\
	& \qquad \qquad \qquad \qquad \qquad \qquad \qquad \qquad \qquad \qquad \qquad - \sin \big( \bfk \cdot (\bfy_\bfp - \bfy_\bfq) \big) \Big] dt
\end{align}
Since the integrals are over a complete period $T=2\pi / \omega$, all terms of the form $\sin(\ldots - 2\omega t \ldots)$ vanish.
Using (\ref{Dynamical1}), this simplifies to:
\begin{equation} \label{EFlux5}
	\psi = \pi \sum_{(\bfp,m) \in \Omega^+} \hat\bfu_m \cdot \sum_{(\bfq,n) \in \Omega^-}  \hat\bfD_{(\bfzero,m)(\bfq-\bfp,n)} \sin \big( \bfk \cdot (\bfy_\bfp - \bfy_\bfq) \big) \hat\bfu_n
\end{equation}

Now consider a nanotube with two generators, i.e. a pure rotation generator with rotation angle $\theta_2=2\pi/N_2$, and a screw generator with the rotation component associated to an angle $-\pi < \theta_1 \le \pi$.
Consider the flow of energy across a surface that divides the OS into $\Omega^- = \{ \bfq: \ q_1 \leq 0 \}$ and $\Omega^+ = \{ \bfq: \ q_1 > 0 \}$, where the first slot in the multi-index corresponds to the screw and the second slot corresponds to the rotation.
Then we can write:
\begin{equation} \label{EFlux7}
	\psi = -\pi \sum_{m,n} \hat\bfu_m \cdot \left[ \sum_{q_2=0}^{N_2-1} \sum_{q_1 \ge 1} q_1 \hat\bfD_{(\bfzero,m)(\bfq,n)} \sin ( k_1 q_1 + k_2 q_2) \right] \hat\bfu_n
\end{equation}
Note that the factor of $q_1$ appears in the sum because, in an OS, various atomic bonds are symmetry-related and therefore the sum need not run over these bonds.

The component $k_2$ of the wave vector corresponds to the pure rotation generator.
Therefore, it takes only the discrete values $k_2=2\pi j / N_2, j=0,\cdots,N_2-1$.

At this point, the model is nominally exact.
Our interest however is in a minimal model that captures the important features.
In terms of energy transport, the wiggles in the phonon spectrum are of primary interest.
We now make extremely harsh simplifying approximations on the nature of interactions, but retain the feature that interactions are non-local in Objective space.
We see that this single feature is sufficient to understand the wiggles.
We assume that (i) interactions are only nearest neighbor in real-space, and (ii) the magnitude of the interactions is the same for all near-neighbors.
Under these assumptions, we search for the values of the wavevector at which $\psi=0$.

Consider a $(6,6)$ nanotube ($N_2 = 6$ in this case).
The bonds that connect $\Omega^+$ and $\Omega^-$ under the assumptions above are $(0,j)$--$(1,j)$ and $(0,j)$--$(1,(j+5)\mod 6)$ for $j=0\ldots 5$.
These are all near-neighbors both in real and Objective space.
Equation (\ref{EFlux7}) specializes to:
\begin{equation} \label{EFlux9}
	6 \left( \sin(k_1) + \sin(5k_2 + k_1) \right) = 0 \Rightarrow 2 \sin(k_1 + 2.5k_2) \cos(2.5k_2) = 0
\end{equation}
Hence, for $k_2=2\pi j/6, j=0,\ldots,5$, the solution is $k_1 = \pi \left(r -\frac{5j}{6} \right), r \in \Z$.
These values match exactly with with the zero-slope points in Fig. \ref{fig:n6m6_2A}.
In addition, there are no wiggles because all interactions are local in Objective space.

Now consider a $(7,6)$ nanotube.
Here $N_2 = 1$ so there is only a single index.
The nearest neighbors of the $0$ unit cell are $6, 7, 13$ and all of these are non-local in Objective space.
Equation \eqref{EFlux7} specializes to:
\begin{equation} \label{EFlux12}
	6\sin(6k_1) + 7\sin(7k_1) + 13\sin(13k_1) =  0
\end{equation}
Numerically solving this gives the wavevectors \\ $\pi \times \{0.000, 0.102, 0.154, 0.206, 0.307, 0.404, 0.463, 0.520, 0.614, 0.703, 0.771, 0.839, 0.921, 1.000 \}$.
This matches extremely well with Fig. \ref{fig:n7m6}, with relative error of the order $10^{-7}$.

We next consider a $(11,9)$, also with $N_2=1$.
The nearest neighbors of the $0$ unit cell are $9,11,20$ and all of these are non-local in Objective space.
Therefore,
\begin{equation} \label{EFlux13}
	9\sin(9k_1) + 11\sin(11k_1) + 20\sin(20k_1) =  0
\end{equation}
Numerically solving for the wavevectors of the wiggles, we find \\ $\pi \times \{ 0.0, 0.066, 0.101, 0.134, 0.199, 0.260, 0.301, 0.342, 0.398, 0.452, \\ 0.501, 0.552, 0.598, 0.643, 0.70, 0.762, 0.798, 0.836, 0.900, 0.967, 1.0 \}$.
This again matches extremely well with the full calculation in Fig. \ref{fig:n11m9}, with relative error on the order of $10^{-8}$.

\begin{figure}[htbp]
	\centering
	\includegraphics[width=160mm]{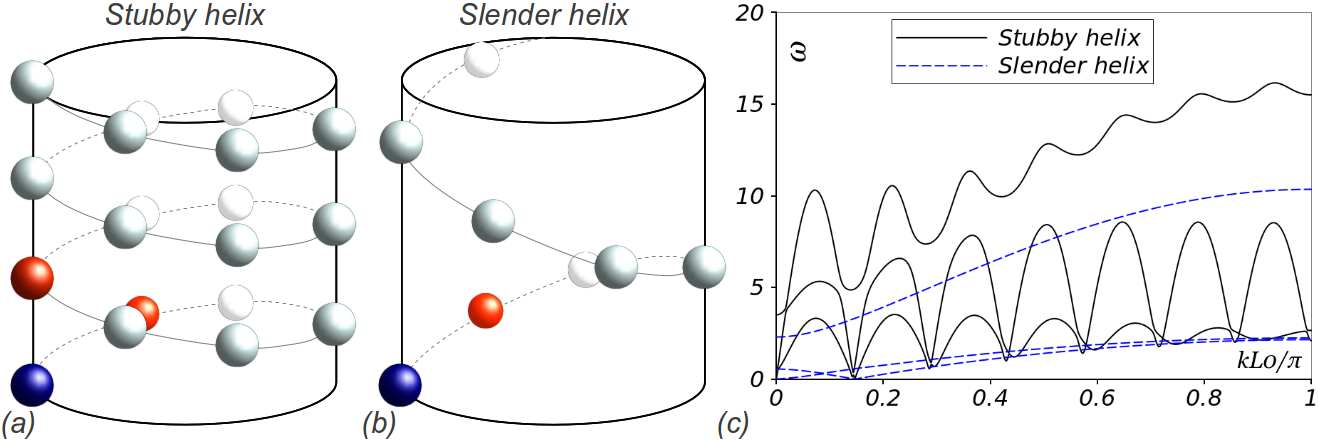}
	\caption{Energy transport in helices with different aspect ratios using model interatomic potentials.  The red atoms are the near-neighbors of the blue atom in real-space.  In the slender helix, there is no long-range interaction and it behaves like a 1D chain.  The stubby helix has long-range interaction and has prominent wiggles in the phonon spectrum.}
	\label{fig:energy-flux-2}
\end{figure}

Finally, we compute the phonon spectra for two nanotubes with different aspect ratios using model interatomic potentials, Fig. \ref{fig:energy-flux-2}.
For the stubby helix, we find prominent wiggles as expected from our model that nonlocal interactions are important.
For the slender helix, it behaves almost like a near-neighbor chain with no long-range interactions as we expect.

\section{Discussion}

We have formulated a method to compute and understand phonon spectra in Objective Structures; this includes a broad class of complex nanostructures.
The use of the OS formulation enables important advantages.
For instance, it is easy to apply complex loads such as torsion.
The framework also enabled us to draw the important conclusion that there is no analog of ``non-rank-one'' instability in nanotubes, i.e., a nanotube that is linearly stable to bending, twisting and elongation individually applied is linearly stable to any superposition of these.

The OS framework also provided a physical interpretation of the computed phonon spectra, thus enabling the construction of the simplified geometric nonlocal model for energy transport.
The primarily geometric nature of the model enables it to be potentially applicable broadly to rod-like helical OS, e.g. biological systems such as DNA.
The simplified model shows an interesting equivalence between curvature and non-locality; a similar equivalence also appears in understanding the kinetics of phase transformations at the atomic level \cite{lu-dayal-inprep}.
In addition, the model predicts well the interplay between axial and helical energy transport mechanisms, in particular the critical points at which these mechanisms destructively interfere have no transport.


\section*{Acknowledgments}

Amin Aghaei and Kaushik Dayal thank AFOSR Computational Mathematics (FA9550-09-1-0393) and AFOSR Young Investigator Program (FA9550-12-1-0350) for financial support.
Kaushik Dayal also acknowledges support from NSF Dynamical Systems (0926579), NSF Mechanics of Materials (CAREER-1150002) and ARO Solid Mechanics (W911NF-10-1-0140).
Ryan S.\ Elliott acknowledges support from NSF CMMI-0746628 (Dr.\ George Hazelnut, Program Director) and The University of Minnesota Supercomputing Institute.
This work was also supported in part by the NSF through TeraGrid resources provided by Pittsburgh Supercomputing Center.
Kaushik Dayal thanks the Hausdorff Research Institute for Mathematics at the University of Bonn for hospitality.
We thank Richard D. James for useful discussions.


\appendix
\section{Relation between Objective group generators and carbon nanotube geometry}
\label{SWNT-parameters}

Consider a carbon nanotube with chiral indices $(m,n)$ and axis $\bfe$ centered at the origin.
Following \cite{dayal-elliott-james-formulas}, we have the following relation between the group generators and the geometry of the carbon nanotube when we use a 2-atom unit cell:
\begin{equation}
	 \label{NT_g1g2}
	\begin{split}
		g_1 & = (\bfR_{\theta_1}|\mathbf 0), \quad \bfR_{\theta_1} \bfe = \bfe, \quad 0<\theta_1=\frac{2\pi \min \left(|p|,|q|\right)}{\GCD(n,m)} \leq 2\pi \\
		g_2 &=(\bfR_{\theta_2}|\kappa_2 \bfe), \quad \bfR_{\theta_2}\bfe = \bfe, \quad \theta_2=\pi \frac{p(2n+m)+q(n+2m)}{n^2+m^2+nm}, \quad \kappa_2=\frac{3l_0 \GCD(m,n)}{2 \sqrt{n^2+m^2+ n m}} \\
	\end{split}
\end{equation}
$\bfR_\theta$ is a rotation matrix with axis coinciding with $\bfe$ and rotation angle $\theta$.
The quantity $l_0 = 0.142$nm is the bond length of the graphene sheet before rolling.
The integers $p$ and $q$ satisfy $pm-qn=\GCD (m,n)$, where $\GCD (m,n)$ is the greatest common divisor of $m$ and $n$.

The radius of the nanotube is $r= \frac{l_0}{2\pi} \sqrt{3(n^2+m^2+nm)}$ and the positions of the atoms in the unit cell are:
\begin{align}
	\bfx_{(0,0),1} = & r \bfe_1  \notag  \\
	\bfx_{(0,0),2} = & r \cos \left[ \frac{\pi (n+m)}{n^2+m^2+nm} \right] \bfe_1 + r \sin \left[ \frac{\pi (n+m)}{n^2+m^2+nm} \right] \bfe_2 + \frac{l_0 (m-n)}{2 \sqrt{n^2+m^2+nm}} \bfe
\end{align}
where $(\bfe, \bfe_1, \bfe_2)$ are orthonormal.

Note that if $m$ and $n$ are relatively prime, i.e. $\GCD(m,n)=1$, then $\theta_1 = 0$ and $g_1$ reduces to the identity.


\section{The Discrete Fourier Transform and Block-Diagonalization of Hessians of Periodic Crystals}
 \label{DFT}

In this appendix, we rewrite the standard Discrete Fourier Transform (DFT) in the notation of matrices and apply this to block-diagonalize the Hessian of a periodic crystal.
This enables a conceptual understanding of the relation between block-diagonalization in crystals and OS.


\subsection{Discrete Fourier Transform and its Properties}

Consider a space with coordinate $\bfy$.
Define the points $\bfy_\bfp$:
\begin{align} \label{XandK-1}
	\bfy_\bfp= & \bfy_{(p_1,p_2,p_3)}=p_1 \bfa_1 + p_2 \bfa_2 + p_3 \bfa_3
\end{align}
where $\bfa_1, \bfa_2, \bfa_3$ is a basis for the space.

Consider a family of periodic functions $\hat\bfu_{\bfp,l}$ that are defined at $\bfy_\bfp$, with the family indexed by $l$.
Physically, these correspond to the displacement of the atom $l$ in the unit cell at $\bfy_\bfp$.
From the periodicity, it follows that $\hat\bfu_{\bfp+\bfN,l}=\hat\bfu_{\bfp,l}$, where $\bfN:=(N_1, N_2, N_3)$ defines the periodicity.

Define the reciprocal basis through $\bfa_\alpha \cdot \bfb_\beta= \delta_{\alpha \beta}$ and the wave vectors as
\begin{align} \label{XandK-2}
	\bfk_\bfq= & \bfk_{(q_1,q_2,q_3)}=\frac{2\pi q_1}{N_1} \bfb_1 + \frac{2\pi q_2}{N_2} \bfb_2 + \frac{2\pi q_3}{N_3} \bfb_3
\end{align}
It follows that $\bfk_\bfq \cdot \bfy_\bfp= \frac{2\pi}{N_1} p_1 q_1 + \frac{2\pi}{N_2} p_2 q_2 + \frac{2\pi}{N_3} p_3 q_3$, or in one-dimension $\bfk_q \cdot \bfy_p= \frac{2\pi}{N} pq$.

These imply that $\bfk_\bfq \cdot \bfy_\bfp = \bfk_\bfp \cdot \bfy_\bfq$.
It also follows that $\exp[i\bfk_\bfq \cdot \bfy_\bfzero]=\exp[i\bfk_\bfq \cdot \bfy_\bfN]=1$ and that $\sum_{j=0}^{N-1} \exp[i\bfk_p \cdot \bfy_j] \exp[-i\bfk_q \cdot \bfy_j] = N \delta_{pq}$.
As we see below, $\exp[i \bfk_\bfp \cdot \bfy_\bfq]$ are the components of the basis vectors of the eigenspace of a circulant matrix, and are closely related to the eigenspace for a block-circulant matrix.

The DFT is defined as
\begin{equation*}
	\tilde u_{\bfp,l}^\alpha= \frac{1}{\sqrt N} \sum_{\bfq} \exp [i\bfk_\bfp \cdot \bfy_{\bfq}] \hat u_{\bfq,l}^\alpha
	\Leftrightarrow \tilde \bfu= \bfF \hat \bfu
\end{equation*}
where $N=N_1N_2N_3$ and $\bfF$ is the DFT matrix which is independent of $l$ and $\alpha$.  The inverse DFT is
\begin{equation} \label{InvFouri}
	\hat u_{\bfp,l}^\alpha=\frac{1}{\sqrt N} \sum_{\bfq} \exp [-i\bfk_\bfq \cdot \bfy_{\bfp}] \tilde u_{\bfq,l}^\alpha 
	\Leftrightarrow \hat \bfu= \bfF^{-1} \tilde \bfu
\end{equation}

This enables us to now represent the standard DFT in terms of matrix notation.
For simplicity, consider a one-dimensional problem, i.e. $\bfp=(p,0,0)$ and $\bfq=(q,0,0)$, and $p$ and $q$ run over the integers between $0$ and $N-1$.
The matrix $\bfF$ can be expressed as
\begin{equation}
	\label{eqn:Fourier-matrix}
	\bfF=
		\begin{bmatrix}
			[\bfF_{00}]     & [\bfF_{01}]     & \cdots & [\bfF_{0(N-1)}]     \\
			[\bfF_{10}]     & [\bfF_{11}]     & \cdots & [\bfF_{1(N-1)}]     \\
			\vdots          & \vdots          & \ddots & \vdots              \\
			[\bfF_{(N-1)0}] & [\bfF_{(N-1)1}] & \cdots & [\bfF_{(N-1)(N-1)}] \\
	     \end{bmatrix}
\end{equation}
and each sub-matrix of $\bfF$ is a $3M \times 3M$ matrix and defined as $[\bfF_{pq}]_{mn}=\frac{1}{\sqrt N} \exp[i \bfk_p \cdot \bfy_q] \delta_{mn}$.

The inverse of the Fourier transform matrix $\bfF^{-1}$ is defined as
\begin{equation}
	\bfF^{-1}=
		\begin{bmatrix}
			[\bfF_{00}^{-1}]     & [\bfF_{01}^{-1}]     & \cdots & [\bfF_{0(N-1)}^{-1}]     \\
			[\bfF_{10}^{-1}]     & [\bfF_{11}^{-1}]     & \cdots & [\bfF_{1(N-1)}^{-1}]     \\
			\vdots               & \vdots               & \ddots & \vdots                   \\
			[\bfF_{(N-1)0}^{-1}] & [\bfF_{(N-1)1}^{-1}] & \cdots & [\bfF_{(N-1)(N-1)}^{-1}] \\
	     \end{bmatrix}
\end{equation}
where $[\bfF_{pq}^{-1}]_{mn}=\frac{1}{\sqrt N} \exp[-i \bfk_q \cdot \bfy_p] \delta_{mn}$.

With these definitions, $\bfF^{-1}=\bfF^\dag$, where $^\dag$ represents the adjoint, and therefore $\bfF$ is unitary.


\subsection{Block-Diagonalization of  a Block-Circulant Matrix using the Discrete Fourier Transform}

Consider a block-circulant matrix, i.e. $\bfA_\mathbf{pq}=\bfA_\mathbf{0(q-p)}$, as arises in periodic crystals.
The DFT provides the block-diagonal matrix $\tilde\bfA = \bfF\bfA\bfF^{-1}$.
\begin{align}
	\tilde\bfA_\mathbf{pq}
	& = \sum_\bfm \sum_\bfn \bfF_\mathbf{pm} \bfA_\mathbf{mn} \bfF_\mathbf{nq}^{-1} \notag \\
	& = \sum_\bfm \sum_\bfn \big( \frac{1}{\sqrt N} \exp[i \bfk_\bfp \cdot \bfy_\bfm] \bfI \big) \bfA_\mathbf{mn} \big( \frac{1}{\sqrt N} \exp[-i \bfk_\bfq \cdot \bfy_\bfn] \bfI \big) \notag \\
	& = \frac{1}{N} \sum_\bfm \sum_\bfn \exp \big[i \bfy_\bfm \cdot \big(\bfk_\bfp - \bfk_\bfq \big) \big] \exp \big[-i \bfk_\bfq \cdot \big(\bfy_\bfn - \bfy_\bfm \big) \big] \bfA_\mathbf{mn} \notag \\
	& = \frac{1}{N} \sum_\bfm \sum_\bfn \exp \big[i \bfy_\bfm \cdot \big(\bfk_\bfp - \bfk_\bfq \big) \big] \exp \big[-i \bfk_\bfq \cdot \big(\bfy_\bfn - \bfy_\bfm \big) \big] \bfA_\mathbf{0(n-m)}
\end{align}
Relabeling $\bfr=\bfn-\bfm$, we can write $\bfy_\bfr=\bfy_\bfn-\bfy_\bfm$.
\begin{align} \label{BlockDiag}
	\tilde\bfA_{\bfp \bfq} 
	& = \frac{1}{N} \sum_\bfm \sum_\bfr \exp \big[i \bfy_\bfm \cdot \big(\bfk_\bfp - \bfk_\bfq \big) \big] \exp \big[-i \bfk_\bfq \cdot \bfy_\bfr \big] \bfA_\mathbf{0r} \notag \\
	& =  \sum_\bfr \exp \big[-i \bfk_\bfq \cdot \bfy_\bfr \big] \bfA_\mathbf{0r} \delta_\mathbf{pq}
\end{align}
using that $\sum_{j=0}^{N-1} \exp[i\bfk_p \cdot \bfy_j] \exp[-i\bfk_q \cdot \bfy_j] = N \delta_{pq}$.
Therefore, $\tilde\bfA$ is block-diagonal.
Further, since  $\bfF \bfF^\dag = \bfI$, we have
\begin{equation}
	\bfA \bfw = \lambda \bfw
	\Rightarrow
	\bfF \bfA \bfF^\dag \bfF \bfw =  \lambda \bfF \bfw 
	\Rightarrow
	\tilde \bfA \tilde\bfw = \lambda \tilde \bfw
\end{equation}

Consider the specific case of the eigenvalue problem \eqref{EqMotion4} in a periodic crystal where $\hat \bfH$ is a block-circulant matrix.
From (\ref{BlockDiag}), $\tilde\bfH$ is block-diagonal and consequently the eigenvalue problem can be expressed as ${\left(\omega^2 \right)}^{[\bfp]} \tilde\bfu_\bfp = \tilde\bfH_\mathbf{pp} \tilde\bfu_\bfp$, where $\tilde\bfH_\mathbf{pp}= \sum_\bfr \exp \big[-i \bfk_\bfp \cdot \bfy_\bfr \big] \hat\bfH_\mathbf{0r}$.

For each $\bfp$, corresponding to a specific wave-vector $\bfk$ in (\ref{XandK-1},\ref{XandK-2}), we obtain $3M$ eigenvalues and eigenvectors corresponding to the different phonon branches.
Denote each branch by $\nu=1,\cdots,3M$, so we can write ${\left(\omega^2 \right)}^{[\bfp,\nu]} \tilde\bfu_\bfp^{[\nu]} = \tilde\bfH_\mathbf{pp} \tilde\bfu_\bfp^{[\nu]}$.
In real-space, the atomic displacements corresponding to the normal mode with wave vector $\bfk_\bfp$ and $\nu$-th branch is  $\hat\bfu_{\bfq,l}^{[\bfp,\nu]} = \frac{1}{\sqrt{N}} \exp[-i \bfk_\bfp \cdot \bfy_\bfq]  \tilde\bfu_{\bfp,l}^{[\nu]}$.


\bibliographystyle{amsalpha}
\bibliography{References}

\end{document}